\documentclass[11pt,article]{elsarticle}

\usepackage{lineno,hyperref}

\usepackage[lmargin=.75in,rmargin=.75in,tmargin=1.in,bmargin=1in]{geometry}
\usepackage{amsmath}
\usepackage{amsthm}
\usepackage{amsfonts}
\usepackage{amssymb}
\usepackage{graphicx}
\usepackage{multicol}
\usepackage{wrapfig}
\usepackage{setspace}
\usepackage{epstopdf}
\usepackage{graphicx}
\usepackage{caption}
\usepackage{subcaption}
\usepackage{algorithm}
\usepackage{color}
\usepackage[noend]{algpseudocode}

\usepackage{framed} 
\usepackage{multicol} 

\immediate\write18{%
makeindex -s nomencl.ist -o \jobname.nls -t \jobname.nlg \jobname.nlo%
}
 
\usepackage{nomencl} 
\makenomenclature
\renewcommand*\nompreamble{\begin{multicols}{2}}
\renewcommand*\nompostamble{\end{multicols}}

\newcommand{\Q}{\mathbf{Q}}
\newcommand{\F}{\mathbf{F}}
\newcommand{\T}{\mathbf{T}}
\definecolor{darkred}{rgb}{.7,0,0}

\journal{Arxiv}

\begin{document}

\begin{frontmatter}

\title{Bayesian inferences of the thermal properties of a wall using temperature and heat flux measurements}

\author[NU]{Marco Iglesias}
\ead{marco.iglesias@nottingham.ac.uk}
\author[KAUST]{Zaid Sawlan \corref{corrauthor}}
\cortext[corrauthor]{Corresponding author}
\ead{zaid.sawlan@kaust.edu.sa}
\author[KAUST,Ur]{Marco Scavino}
\ead{marco.scavino@kaust.edu.sa}
\author[KAUST]{Ra\'ul Tempone}
\ead{raul.tempone@kaust.edu.sa}
\author[NUA]{Christopher Wood}
\ead{christopher.wood@nottingham.ac.uk}

\address[NU]{School of Mathematical Sciences, University of Nottingham, Nottingham, UK}
\address[KAUST]{King Abdullah University of Science and Technology, CEMSE, Thuwal, Saudi Arabia}
\address[NUA]{Department of Architecture and Built Environment, University of Nottingham, Nottingham, UK}
\address[Ur]{Instituto de Estad\'{\i}stica (IESTA), Universidad de la Rep\'ublica, Montevideo, Uruguay}

\begin{abstract}
The assessment of the thermal properties of walls is essential for accurate building energy simulations that are needed to make effective energy-saving policies. These properties are usually investigated through in-situ measurements of temperature and heat flux over extended time periods. The one-dimensional heat equation with unknown Dirichlet boundary conditions is used to model the heat transfer process through the wall. In [F. Ruggeri, Z. Sawlan, M. Scavino, R. Tempone, A hierarchical Bayesian setting for an inverse problem in linear parabolic PDEs with noisy boundary conditions, Bayesian Analysis 12 (2) (2017) 407--433], it was assessed the uncertainty about the thermal diffusivity parameter using different synthetic data sets. In this work, we adapt this methodology to an experimental study conducted in an environmental chamber, with measurements recorded every minute from temperature probes and heat flux sensors placed on both sides of a solid brick wall over a five-day period. The observed time series are locally averaged, according to a smoothing procedure determined by the solution of a criterion function optimization problem, to fit the required set of noise model assumptions. Therefore, after preprocessing, we can reasonably assume that the temperature and the heat flux measurements have stationary Gaussian noise and we can avoid working with full covariance matrices. The results show that our technique reduces the bias error of the estimated parameters when compared to other approaches. Finally, we compute the information gain under two experimental setups to recommend how the user can efficiently determine the duration of the measurement campaign and the range of the external temperature oscillation.
\end{abstract}



\end{frontmatter}

\section{Introduction}
\label{sec_intro}

Concerns about climate change and the effects of greenhouse gases have led to international targets for reducing carbon emissions \cite{Kyoto2014ratification,climate2008change}. One substantial source of carbon emissions is the built environment, which accounts for approximately one-third of global energy consumption \cite{iea2016ee}. For example, approximately 40\% of national energy consumption in the UK is from the building sector. Reduction in carbon emissions from the built environment is, therefore, vital to meeting carbon reduction targets. Carbon emissions from buildings can be considerably reduced through large-scale policies that seek to limit energy demand for space heating and cooling  \cite{iea2016ee}. Accurate predictions of building performance and energy demands are essential to the success of such policies. Specifically, computer simulations of heat loss from buildings are necessary to assess the effectiveness of energy-saving strategies such as retrofit interventions \cite{hong2006impact}. However, recent works \cite{cesaratto2013measuring,asdrubali2014evaluating,solid} have shown that standard computer simulations of building performance may be unreliable due to inaccuracies from poorly characterized building structures including walls. Energy-saving measures based on inaccurate predictions of building performance may be economically ineffective.

Uncertainty in the thermal properties of walls is a primary source of inaccuracy in predictions of energy demand in buildings \cite{solid,Wit}. The heat capacitance and thermal conductance (resistance) of walls are used in standard heat transfer models as parameters for building performance simulations. Since these parameters of existing buildings are often unknown, the corresponding inputs in building simulations are typically obtained by visual inspection and tabulated values. In most cases, these values do not provide accurate characterizations of the walls of the building under consideration. 

\begin{table}[!t]
\begin{framed}
\nomenclature[11]{$R$}{Thermal resistance or R-value, $m^2K/W$}
\nomenclature[12]{$\rho$}{Density of the material, $kg/m^3$}
\nomenclature[13]{$c_p$}{Specific heat capacity, $J/kg K$}
\nomenclature[14]{$k$}{Thermal conductivity, $W/m K$}
\nomenclature[15]{$\rho C$}{Heat capacity of unit area, $J/m^2K$}
\nomenclature[16]{$L$}{Wall thickness, $m$}
\nomenclature[17]{$T(x,t)$}{Temperature at position $x$ and time $t$, $^{\circ}C$}
\nomenclature[18]{$T_0$}{Initial temperature, $^{\circ}C$}
\nomenclature[19]{$T_{int}(t), T_{ext}(t)$}{Surface temperatures of internal and external walls at time $t$, $^{\circ}C$}
\nomenclature[21]{$\mathbf{T}_{int}, \mathbf{T}_{ext}$}{Surface temperatures of internal and external walls at times $t_0, \ldots, t_N$}
\nomenclature[22]{$\mathbf{Y}_{int}, \mathbf{Y}_{ext}$}{Surface temperatures of internal and external walls at times $t_0, \ldots, t_N$}
\nomenclature[23]{$\boldsymbol{\mu}_{int}, \boldsymbol{\mu}_{ext}$}{Smoothed surface temperature measurements of internal and external walls at times $t_0, \ldots, t_N$}
\nomenclature[24]{$C_{int}, C_{ext}$}{Surface temperature noise covariance matrices of internal and external walls}
\nomenclature[25]{$F_{int}(t), F_{ext}(t)$}{Heat fluxes of internal and external walls at time $t$, $W/m^2$}
\nomenclature[26]{$\mathbf{F}_{int}, \mathbf{F}_{ext}$}{Heat fluxes of internal and external walls at times $t_0, \ldots, t_N$}
\nomenclature[27]{$\mathbf{Q}_{int}, \mathbf{Q}_{ext}$}{Heat flux measurements of internal and external walls at times $t_0, \ldots, t_N$}
\nomenclature[28]{$\Sigma_{int}, \Sigma_{ext}$}{Heat flux noise covariance matrices of internal and external walls}
\nomenclature[29]{$\theta$}{Unknown parameters $(R, \rho C, \tau_0)$}
\nomenclature[31]{$\lambda$}{Smoothing parameter}
\nomenclature[34]{$\pi$}{Probability density function}
\nomenclature[35]{$\pi_p$}{Prior probability density function}
\nomenclature[36]{$\mathcal{L}$}{Likelihood function}
\nomenclature[37]{$\xi$}{Experimental setup}
\nomenclature[38]{$D_{KL}$}{Information gain}
\printnomenclature
\end{framed}
\end{table}

The thermal properties of walls can be inferred from in-situ measurements of temperature and heat flux \cite{bidd,luo,solid}. More specifically, the surface temperatures of internal and external walls denoted as $\{T_{int}^{i}\}_{i=1}^{N}$ and $\{T_{ext}^{i}\}_{i=1}^{N}$, are measured at a specified location over time. In addition, the heat flux through the wall, $\{q^{i}\}_{i=1}^{N}$, is also measured at $N$ equispaced time points. ISO 9869:2014 \cite{ISO869:2014} outlines a simple averaging procedure to determine the thermal transmittance (U-value) from in-situ measurements. With this approach, the R-value (i.e., the inverse of the U-value) is computed directly by
\begin{equation*}
\label{average_method}
R=\frac{\sum_{i=1}^{N}(T_{int}^{i} - T^{i}_{ext})}{\sum_{i=1}^{N} q^i} \,.
\end{equation*}
Since the averaging procedure assumes that the thermal mass of the wall is zero or almost zero, the accuracy of the estimate of the U-value will require measurements collected over an extended period of time (often longer than two weeks) \cite{bidd,ISO869:2014}. More importantly, the averaging method does not provide a statistical framework that accounts for either the uncertainty in the thermal properties or errors in the measurements. As a result, this method fails to provide a proper quantification of the uncertainty in the estimated U-value of the wall.

Recent work has proposed the use of statistical approaches to infer thermal properties from in-situ measurements of temperature and heat flux with simplified heat transfer models. In particular, a standard Bayesian inference has recently been proposed \cite{bidd} to estimate the thermal properties of walls under the assumption that the heat dynamics of the wall can be described with a simple lumped-mass resistance-capacitance (RC) model. In contrast to the averaging method, the approach in \cite{bidd} employs an RC network whose parameters include the thermal conductivity and the heat capacity. This standard Bayesian methodology suggests that these thermal properties can be inferred from in-situ measurements based on relatively shorter measurement campaigns than the ones required by the averaging method. While other non-Bayesian statistical methods for estimating thermal properties have been proposed \cite{Gutschker2008163}, \cite{bidd} provides substantial insight into the advantages of using Bayesian inference in building models and provides a motivation for further developments. 

Our present work is a special application of parameter estimation for partial differential equations \cite{xun2013parameter,muller2002fitting}. In particular, we develop and implement the hierarchical Bayesian approach introduced in \cite{Sawlan}. Related works on Bayesian inference used for different applications can be found in the literature (see, for example, \cite{gnanasekaran2011bayesian,parthasarathy2008estimation,toivanen2012simultaneous}). A general Bayesian formulation of inverse problems in heat transfer is also available in \cite{kaipio2011bayesian,orlande2012inverse}. However, we address a problem where the boundary conditions can not be assumed known and a Bayesian approach based on the full likelihood function will not be recommended. Instead, the strength of our approach is to construct data-driven Gaussian priors \cite{Sawlan}, treating the boundary conditions as nuisance parameters to be marginalized out, to develop a quick and applicable Bayesian assessment of the parameters of interest. In \cite{Sawlan}, this approach was implemented to infer the thermal diffusivity parameter using synthetic temperature measurements in the interior and boundary of the domain. Here, we adapt the methodology to deal with temperature and heat flux measurements that are only available on the boundaries. We provide the maximum likelihood estimate (MLE) and the posterior distributions of the unknown parameters. Under the specification of independent uniform priors for the parameters of interest, we first use the Laplace method to produce fast estimates of their posterior distributions \cite{Ghosh,rue2009approximate}. Then, we apply a Markov chain Monte Carlo (MCMC) sampling algorithm to assess the accuracy of the approximations obtained via Laplace method. The MCMC simulations, for this problem, support the employment of Laplace method to speed up the computations and estimate information gain values.  

Most existing methodologies for inferring thermal properties \cite{bidd} use forward models that can be derived from simplified coarse-grid approximations (often with 2 or 3 spatial nodes) of the heat equation that describes heat transfer through a wall. These simplified models are often used for the sake of computational expediency in the parameter identification process. However, such simplifications introduce intrinsic modeling errors that may, in turn, result in biased and potentially inaccurate estimated parameters. Alternatively, we use a heat equation with unknown Dirichlet boundary conditions to model the interior temperature of the wall, and we provide a convergence analysis to assess the effect of the discretization error in the Bayesian estimates of the thermal properties. We show that the proposed technique is robust under grid refinement and is, therefore, suitable for any discretization that we may choose according to the computational resources at hand.

In-situ temperature measurements of a wall are often used as boundary conditions for a forward heat transfer model of the wall. In the Bayesian approach in \cite{bidd} the inference of the thermal properties is made by inverting heat flux observations while using measurements of temperatures for the forward heat transfer model. Only the heat flux measurements are used to construct the likelihood function. Temperature measurements are assumed ``exact'' and are utilized as boundary conditions for the forward RC model in the Bayesian framework for inferring thermal properties. In contrast, our hierarchical approach accounts for uncertainty in the temperature measurements by treating the nuisance boundary conditions as random functions modeled by Gaussian distributions. As we show in Section \ref{sec3}, failing to account for the uncertainty in these measurements can result in biased and inaccurate estimates of the inferred properties. We also compare our results with those obtained by applying a similar approach to the one in \cite{bidd}, in which smooth spline fits of the temperature measurements are used as exact boundary conditions.

Standard protocols based on asymptotic assumptions \cite{ISO869:2014} require long measurement campaigns during winter to reduce the dynamic effect of the capacitance of the wall. However, \cite{bidd} suggests that shorter measurements campaigns may provide similar estimates of inferred parameters to the ones from longer measurement campaigns. Hence, we use the proposed hierarchical Bayesian framework to investigate the effects of the duration and the conditions (i.e., measurement cycle) of the measurement campaign. To this end, we estimate, by Laplace method, the information gain \cite{Kull, Quan} to quantify the duration of the measurement campaign and the corresponding cycle. The proposed approach can then be used to design cost-effective measurement campaigns.

The rest of this paper is organized as follows. In Section \ref{sec2}, we derive a simulation model of the heat flow process through a wall using the heat equation. Based on simple assumptions, we reduce the model to a one-dimensional heat equation with unknown Dirichlet boundary conditions and write the modeled heat flux as a linear function of the boundary conditions. The Bayesian approach is then introduced; we construct the joint likelihood by assuming Gaussian noise in the heat flux measurements. We also assume that the nuisance boundary conditions are random functions modeled by Gaussian distributions. Under these assumptions, the marginalization of the boundary conditions can be performed analytically. Section \ref{sec3} includes the description of the experiment conducted at the Nottingham University Innovation Park to collect temperature and heat flux measurements from both sides of a brick wall. Moreover, smoothing time series techniques are applied to the real data to assess the relevance of the measurement error. Example 1 shows how the thermal properties of the wall can be estimated when a deterministic approximation of the nuisance boundary conditions is used. The numerical results of the marginalization technique are then presented in Example 2, where we also study the convergence of the ML estimates of the model parameters and estimate bivariate posterior distributions. In Subsection \ref{Robust}, we analyze the robustness of our Bayesian approach by means of bootstrap resampling methods. Finally, we compute the information gain about the model parameters under different experimental setups in Section \ref{InfGain}. 

\section{Methodology}
\label{sec2}

In this Section, we describe the forward and inverse methodologies used to characterize the thermal properties of walls. We also introduce the heat transfer (forward) model that we combine with the hierarchical Bayesian methodology introduced in Subsection \ref{inverse}.

\subsection{The Forward model}\label{forward}

The existing forward models used to infer the thermal properties of walls are based on simplified heat transfer models \cite{bidd}. Inferring parameters (i.e., thermal properties) from such simplified models can be done in a computationally affordable fashion through standard identification/inference techniques. However, as we stated earlier, this oversimplification of the heat transfer process might lead to modeling errors that are often not incorporated into standard inference approaches. In this Section, we propose a realistic mathematical model to simulate the heat transfer process through a wall using an initial/boundary value formulation for the heat equation. 

\subsubsection{Heat equation}

The heat transport process inside a wall is modeled, in general, by a three-dimensional heat equation on a rectangular prism, $\Omega$. The initial/boundary value problem for the heat equation along the period $[0, t_N]$, with initial wall temperature $T_0$ and temperature profile $T_s$ at the boundary $\Gamma$, is given by:

\vspace{0.1in}

\begin{minipage}{.60\textwidth}
\begin{eqnarray*}
\rho c_p \frac{\partial T}{\partial t} &=& \nabla \cdot ( k(\bold{x}) \nabla T),\,  \bold{x} \in \Omega,\,  t \in [0,t_{N}] \\
T(\bold{x},t) &=& T_{s}(\bold{x},t),\,  \bold{x} \in \Gamma,\,  t \in [0,t_{N}] \\
T(\bold{x},0) &=& T_0(\bold{x}),\, \bold{x} \in \Omega
\end{eqnarray*}
where $\rho, c_p$ and $k$ denote the density of the material, the specific heat capacity and the thermal conductivity, respectively.
\end{minipage}
~~
\begin{minipage}{.40\textwidth}
 \includegraphics[width =0.45\textwidth]{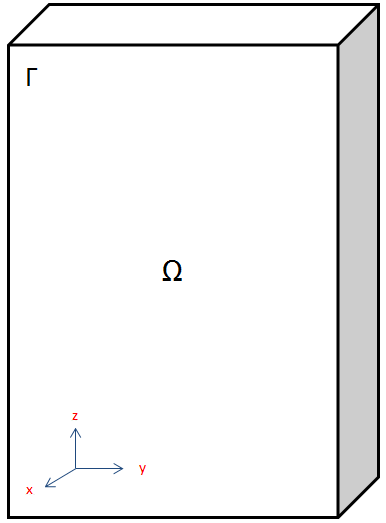}
\end{minipage}

\vspace{0.1in}

Here, we consider the specific situation where the wall is surrounded by insulation materials and its thickness is less than its length and width. We, therefore, assume that the wall temperature varies only along the thickness dimension denoted by $x$. Also, given that the wall is solid, the thermal conductivity is assumed to be constant, $k(x):=k$. As a consequence, our simulation model consists of a one-dimensional heat equation with Dirichlet boundary conditions. From the solution of the heat equation, we then define the heat flux functions, $F_{int}$ and $F_{ext}$, which correspond to the model predictions of heat fluxes at the boundary. The model under consideration takes the following form:
\begin{equation}
\begin{cases}
\rho c_p \frac{\partial T}{\partial t} =  \frac{\partial}{\partial x} \left( k \, \frac{\partial T}{\partial x} \right),\,  x \in (0, L),\,  t \in [0,t_{N}] \\
T(0,t) = T_{int}(t),\,  T(L,t) = T_{ext}(t),\,  t \in [0,t_{N}] \\
T(x,0) = T_0(x),\, x \in (0, L), \\
F_{int}(t) = - k \frac{\partial T}{\partial x}|_{x=0}, \\
F_{ext}(t) = - k \frac{\partial T}{\partial x}|_{x=L},
\end{cases}
\label{heateq}
\end{equation}
where $L$ is the wall thickness, $T_{int}$ and $T_{ext}$ are the internal and external wall surface temperatures, respectively. In our implementation, we replace the thermal conductivity, $k$, with $\frac{L}{R}$ and replace $\rho c_p$, with $\frac{\rho C}{L}$. Then, we estimate the thermal resistance, $R$, and the heat capacity per unit area, $\rho C$, which are properties of the wall.

\subsubsection{Numerical approximation} 
\label{NumApp}

In the numerical implementation of the forward heat transfer model, it is simple to show, upon discretization, that the heat flux can be written as a linear function of the initial condition, $T_{0}$, and the boundary conditions, $T_{int}$ and $T_{ext}$. 

Consider the following uniform space-time discretization,
\[ x_0 = 0, x_1 =  \Delta x, \ldots, x_m = m \Delta x, \ldots, x_M = M \Delta x = L \, , \] 
\[ t_0 = 0, t_1 =  \Delta t, \ldots, t_n = n \Delta t, \ldots, t_N = N\Delta t \, , \] 
and define the vectors 
\[  \mathbf{T}_{int} = (T_{int}(t_0), \ldots, T_{int}(t_{N}) )' \, , \mathbf{T}_{ext} = (T_{ext}(t_0), \ldots, T_{ext}(t_{N}) )' \, ,\]
\[ \mathbf{T_0} = (T_{0}(x_1), \ldots, T_{0}(x_{M-1}) )' \, ,\]
\[  \mathbf{F}_{int} = (F_{int}(t_0), \ldots, F_{int}(t_{N}) )' \, ,  \mathbf{F}_{ext} = (F_{ext}(t_0), \ldots, F_{ext}(t_{N}) )' \, .\]

Then, the discretized heat flux can be approximated by a linear function of the initial condition, $\mathbf{T}_{0}$, and the boundary conditions, $\mathbf{T}_{int}$ and $\mathbf{T}_{ext}$:
\begin{eqnarray}
\label{numapp}
\mathbf{F}_{int} &\approx& H \mathbf{T}_{0} + H_{int} \mathbf{T}_{int} + H_{ext} \mathbf{T}_{ext}, \\  
\mathbf{F}_{ext} &\approx& G \mathbf{T}_{0} + G_{int} \mathbf{T}_{int} + G_{ext} \mathbf{T}_{ext}, \label{HFapp}
\end{eqnarray}
where $H,H_{int}, H_{ext}, G, G_{int}$ and $G_{ext}$ are matrices that may depend nonlinearly on the parameters $R$ and $\rho C$, which, in turn, we infer in the next Subsection. Such matrices are explicitly defined in the Appendix (\ref{appA}). In the case in which thermal conductivity is not constant, the previous result can be proved using the finite element method instead of the finite difference method (see \cite{Sawlan}).

In the subsequent analysis, we assume that the initial condition, $T_{0}(x)$ (and its corresponding discretization $\mathbf{T}_{0}$), is well approximated by the piecewise linear function
\begin{equation}\label{model_as}
\begin{cases} 
      T_{int}(0) + 2 \frac{\tau_{0} - T_{int}(0)}{L} x & \textrm{ if $0 < x \leq \frac{L}{2}$} \\
      \tau_0 + 2\frac{T_{ext}(0) - \tau_{0}}{L} (x - \frac{L}{2}) & \textrm{ if $\frac{L}{2} < x < L$} \, ,
   \end{cases}
\end{equation}
where $\tau_0$ is an unknown constant parameter.

The discretized model (\ref{numapp})-(\ref{HFapp}) can be written as 
\begin{eqnarray}
\label{numapp2}
\left[\begin{array}{c}
\mathbf{F}_{int}  \\   
\mathbf{F}_{ext}\end{array}\right]=\mathcal{F}(R, \rho C,\tau_{0},\T_{int},\T_{ext})\, ,
\end{eqnarray}
where $\mathcal{F}$ is a non-linear function that arises from the numerical discretization (\ref{numapp})-(\ref{HFapp}) and the modeling assumption (\ref{model_as}). 

\subsection{The Bayesian inverse problem}\label{inverse}

Assume that we have noisy measurements of the heat fluxes, $\mathbf{F}_{int}$ and $\mathbf{F}_{ext}$, at the observation times, $t_0, \ldots, t_N$. We denote these measurements as $\mathbf{Q}_{int} = \{ Q^0_{int}, ..., Q^{N}_{int} \}$ and $\mathbf{Q}_{ext} = \{ Q^0_{ext}, ..., Q^{N}_{ext}\}$, respectively. Similarly, we assume that noisy measurements of the boundary temperatures, $\mathbf{T}_{int}$ and $\mathbf{T}_{ext}$, are taken at those observation times. These observations are denoted by $\mathbf{Y}_{int} = \{ Y^0_{int}, ..., Y^{N}_{int} \}$ and $\mathbf{Y}_{ext} = \{ Y^0_{ext}, ..., Y^{N}_{ext}\}$, respectively. The objective of the proposed Bayesian methodology is to estimate $\theta = (R, \rho C, \tau_{0})$  given heat flux measurements $(\mathbf{Q}_{int}, \mathbf{Q}_{ext}$) and boundary temperature measurements $(\mathbf{Y}_{int}, \mathbf{Y}_{ext})$. Whereas the parameters $R$ and $\rho C$ are the unknown variables of interest that characterize the thermal properties of the wall, the initial temperature parameter, $\tau_{0}$, is also unknown. It must be inferred alongside $R$ and $\rho C$. 

\subsubsection{The Bayesian approach} \label{bayesapp}

We adopt the Bayesian approach in which the goal is to find the probability distribution of the unknown parameters, $\gamma$, given the data, namely the posterior distribution, $\pi (\gamma| data)$. From Bayes' theorem, the posterior distribution of $\gamma$ is given by
\begin{equation*}
\pi(\gamma |data) \propto  \pi(data | \gamma) \pi_{p}(\gamma),
\end{equation*}
where $\pi(data | \gamma) = \mathcal{L}(\gamma | data)$ is the likelihood function of $\gamma$ and $\pi_p$ is the prior distribution of $\gamma$ \cite{sivia2006}. In the context of the model defined by expression (\ref{numapp2}), we can see that the parameters of the model are $\theta\equiv (R,\rho C, \tau_{0})$ as well as $\mathbf{T}_{int}$ and $\mathbf{T}_{ext}$. Once these parameters are prescribed, expressions (\ref{numapp})-(\ref{HFapp}) determine the model predictions of the heat fluxes. Therefore, the joint posterior distribution of all these parameters is given by
\begin{equation}\label{joint_li}
\pi( \theta ,\mathbf{T}_{int}, \mathbf{T}_{ext}  | \mathbf{Q}_{int}, \mathbf{Q}_{ext} )\propto \pi(\mathbf{Q}_{int}, \mathbf{Q}_{ext} | \theta ,\mathbf{T}_{int}, \mathbf{T}_{ext}  ) \pi_{p}(\theta , \mathbf{T}_{int}, \mathbf{T}_{ext}).
\end{equation}
Note that $\theta$ is the unknown parameter of interest that contains the thermal properties of the wall. 

In contrast, the boundary parameters, $\mathbf{T}_{int}$ and $\mathbf{T}_{ext}$, are related to noisy measurements. Therefore, these nuisance boundary parameters will be marginalized from the joint likelihood. The specification of the prior distributions of these nuisance parameters is driven by the data by means of the boundary temperature measurements $(\mathbf{Y}_{int}, \mathbf{Y}_{ext})$. As a result, the joint likelihood function will be based only on the heat flux measurements. Nevertheless, by incorporating $\mathbf{T}_{int}$ and $\mathbf{T}_{ext}$ as unknown parameters in this hierarchical fashion, we are effectively taking into account the uncertainty in the corresponding temperature measurements $(\mathbf{Y}_{int}, \mathbf{Y}_{ext})$. This is a new contribution relative to \cite{bidd} where the noise in these observations is neglected. 

\subsubsection{Joint likelihood}

To construct the joint likelihood $\pi(\mathbf{Q}_{int}, \mathbf{Q}_{ext} | \theta ,\mathbf{T}_{int}, \mathbf{T}_{ext}  )$, we assume that the noises of the heat flux measurements are independent Gaussian:
\begin{eqnarray*}
\left( \mathbf{Q}_{int} - \mathbf{F}_{int} \right) \Big| _{\{ \theta, \mathbf{T}_{int}, \mathbf{T}_{ext} \}} \sim N(\bold{0}, \Sigma_{int}), \\
\left( \mathbf{Q}_{ext} - \mathbf{F}_{ext} \right) \Big| _{\{ \theta, \mathbf{T}_{int}, \mathbf{T}_{ext} \}} \sim N(\bold{0}, \Sigma_{ext}).
\end{eqnarray*}
The Gaussianity assumption will be satisfied by the data used in the Bayesian analysis in Section \ref{sec3}.

Under the aforementioned assumptions, the joint likelihood function of $\theta, \mathbf{T}_{int}, \mathbf{T}_{ext}$ is given by
\begin{equation}
\label{JL}
\begin{aligned}
& \mathcal{L}( \theta, \mathbf{T}_{int}, \mathbf{T}_{ext},  \Big| \mathbf{Q}_{int}, \mathbf{Q}_{ext} )  \\
&= \frac{1}{(2 \pi)^{N} \sqrt{ |\Sigma_{int}| |\Sigma_{ext}| } } \exp \left\{ -\frac{1}{2} \left( ||\mathbf{Q}_{int} - \mathbf{F}_{int}||_{\Sigma_{int}}^2 + ||\mathbf{Q}_{ext} - \mathbf{F}_{ext}||_{\Sigma_{ext}}^2 \right) \right\} \,.
\end{aligned}
\end{equation}

We emphasize that $\mathbf{T}_{int}$ and $\mathbf{T}_{ext}$ are nuisance parameters that appear in the formulation via the forward model (i.e., the heat equation).  A direct approach to eliminating these parameters is to set $\mathbf{T}_{int}=\mathbf{Y}_{int}$ and $\mathbf{T}_{ext}=\mathbf{Y}_{ext}$  (recall that $\mathbf{Y}_{int}$ and $\mathbf{Y}_{ext}$ are noisy measurements of the boundary temperatures) or to set $\mathbf{T}_{int}=\boldsymbol{\mu}_{int}$ and $\mathbf{T}_{ext}=\boldsymbol{\mu}_{ext}$ where $\boldsymbol{\mu}_{int}$ and $\boldsymbol{\mu}_{ext}$ are smoothed versions of $\mathbf{Y}_{int}$ and $\mathbf{Y}_{ext}$, respectively. This approach is used in \cite{bidd} where temperature measurements are considered to be deterministic boundary conditions of the RC model. Instead, we eliminate the aforementioned nuisance parameters by marginalizing them using data-driven priors \cite{Sawlan}. This marginalization, which we conduct in the subsequent Subsection, enables us to account for the uncertainty in temperature measurements. As we demonstrate in Subsection \ref{Ex2}, the marginalization process removes the bias in the inferred parameters, thereby providing accurate estimates and reliable quantification of their uncertainty. Moreover, rather than the simple RC model used in \cite{bidd}, here we consider more advanced model given by the heat equation introduced earlier.

\subsection{Marginal likelihood}
\label{MargLike}

In this Section, we use temperature measurements to construct the data-driven Gaussian priors. We perform analytical integration to marginalize out the boundary conditions and obtain a marginal likelihood for $\theta$. We assume that the errors related to the nuisance boundary parameters are independent Gaussian with covariances $C_{int,p}$ and $C_{ext,p}$ as follows:
\begin{equation}
\label{temp_Gauss}
 \mathbf{T}_{int} - \boldsymbol{\mu}_{int} \sim N( \bold{0}, C_{int,p} ), \mathbf{T}_{ext} - \boldsymbol{\mu}_{ext} \sim N( \bold{0}, C_{ext,p} ) ,
\end{equation}
where $\boldsymbol{\mu}_{int}$ and $\boldsymbol{\mu}_{ext}$ are smoothing splines constructed from the boundary temperature measurements. 

The marginal likelihood of $\theta$ is given by
\begin{equation}\label{mar_li}
\begin{aligned}
 \mathcal{L}( \theta | \mathbf{Q}_{int}, \mathbf{Q}_{ext}) &\propto |\Lambda_0|^{1/2} |\Lambda_1|^{1/2} \exp \big\{ - \frac{1}{2} U +\frac{1}{2} t'_{int,2} \Lambda_0 t_{int,2} +  \frac{1}{2} t'_{ext,1} \Lambda_1 t_{ext,1} \big\} \, ,
\end{aligned}
\end{equation}
where $\Lambda_0, \Lambda_1, U, t_{int,2}$ and $t_{ext,1}$ are independent of $\mathbf{T}_{int}$ and $\mathbf{T}_{ext}$ and explicitly defined in the Appendix (\ref{appB}). The marginal likelihood (\ref{mar_li}) can now be used in the Bayesian framework, summarized in Subsection \ref{bayesapp}, to compute the posterior distribution, $\pi(\theta \vert \mathbf{Q}_{int},\mathbf{Q}_{ext})$. 

\section{Experimental data and numerical results}
\label{sec3}
In this Section, we apply the proposed Bayesian approach to infer the thermal properties of a wall from in-situ measurements of heat flux and temperature collected under controlled conditions. In Subsection \ref{ex_set_up}, we describe the experimental setup. The results from the Bayesian analysis are presented in Subsections \ref{Ex1} and \ref{Ex2}.

\subsection{Experimental setup}\label{ex_set_up}
Data were collected from an experiment conducted inside an environmental chamber in the Energy Technologies Building, Nottingham University Innovation Park. The chamber consisted of two rooms separated by a $215-$mm thick partition wall. The two rooms had internal dimensions of $3.70 \times 3.50 \times 2.38$ m. The data were collected from a $970 \times 600 \times 215-$mm brick section of the partition wall. Heat flux sensors and temperature probes were placed on both sides of the bricks. 

The heat flux sensors are Hukseflux (HFP01) sensors with an accuracy of $\pm 5\%$. The surface temperatures were measured with platinum resistance sensors (PT100) with an accuracy of $\pm 0.1^{\circ}\mathrm{C}$. The surface temperature was taken directly on the wall next to the heat flux sensor. This instrumental setup was replicated on each side of the wall. All the sensors were connected to a DataTaker DT85 data logger and readings were recorded at 1 minute intervals.

 
According to CIBSE Guide A \cite{cibse2015environmental} (Tables 3.38, 3.47), reference values of $R$ and $\rho C$ for the wall under consideration should be in the range of $[0.279, 0.3839]$ ($m^2 K/W)$ and $[3.01\times 10^5,3.76\times 10^5]$ ($J/m^2 K$), respectively. The temperature in Room 2 fluctuated based on hourly weather data collected from Nottingham city during $8$ to $15$ February 2014.

\begin{figure}[h!]
\centering
\includegraphics[width=0.8\textwidth]{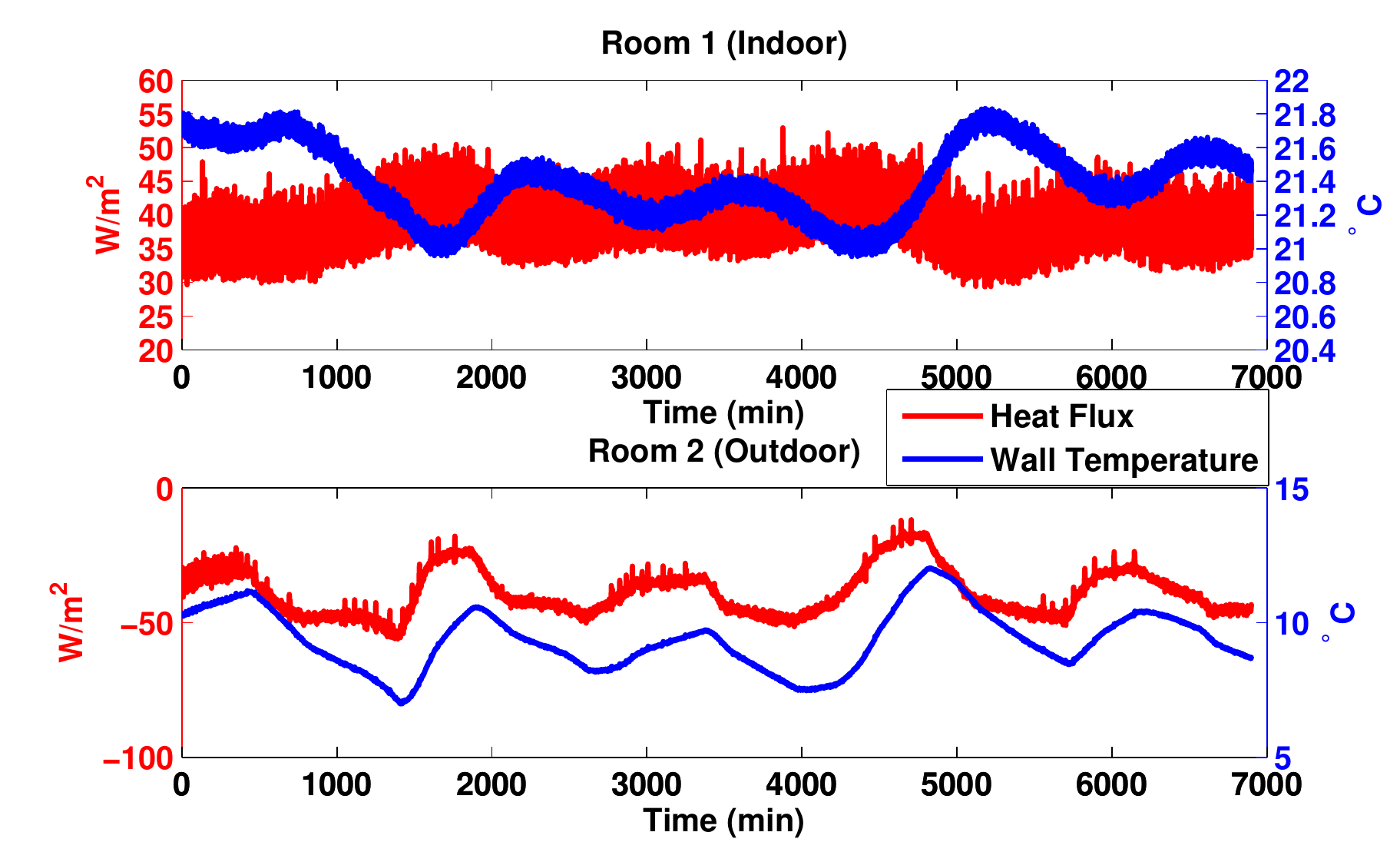}
\caption{Raw temperature and heat flux measurements. Temperature in Room 2 imitates outdoor weather conditions.}
\label{fig1}
\end{figure}

Figure \ref{fig1} shows the temperature and heat flux time series, each consisting of $6,900$ measurements, with a recording interval of one measurement per minute. Clearly, these raw measurements are contaminated by unknown noise. To analyze this noise, we use a smoothing spline method to fit a curve to each time series. This approach is based on the reasonable assumption that the real temperature and heat flux, according to the characteristics of the conducted experiment, vary smoothly over time. The noise is then approximated by the difference between the raw measurements and the smooth values. 

\begin{figure}[h!]
\centering
\includegraphics[width=1.0\textwidth]{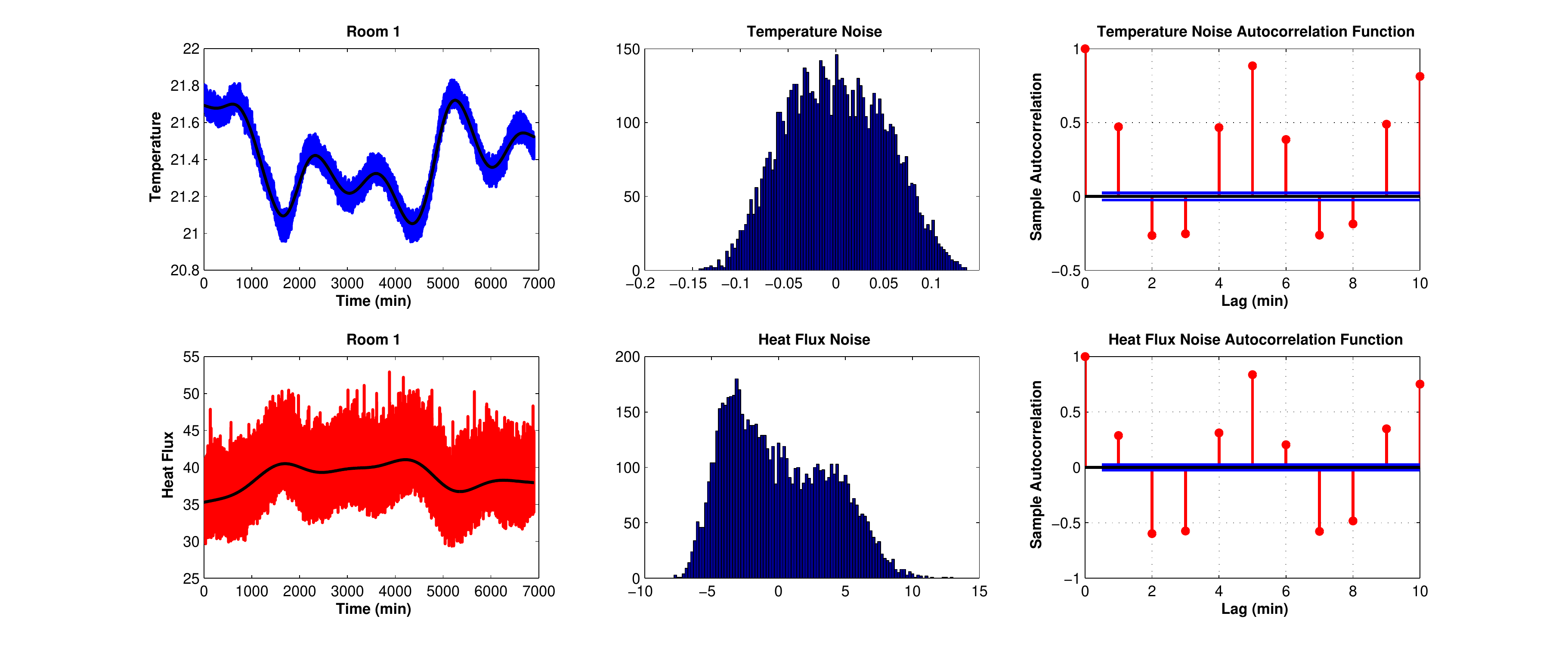}
\caption{Estimated noise of the raw temperature and heat flux measurements in Room 1.}
\label{fig3}
\end{figure}  

\begin{figure}[h!]
\centering
\includegraphics[width=1.0\textwidth]{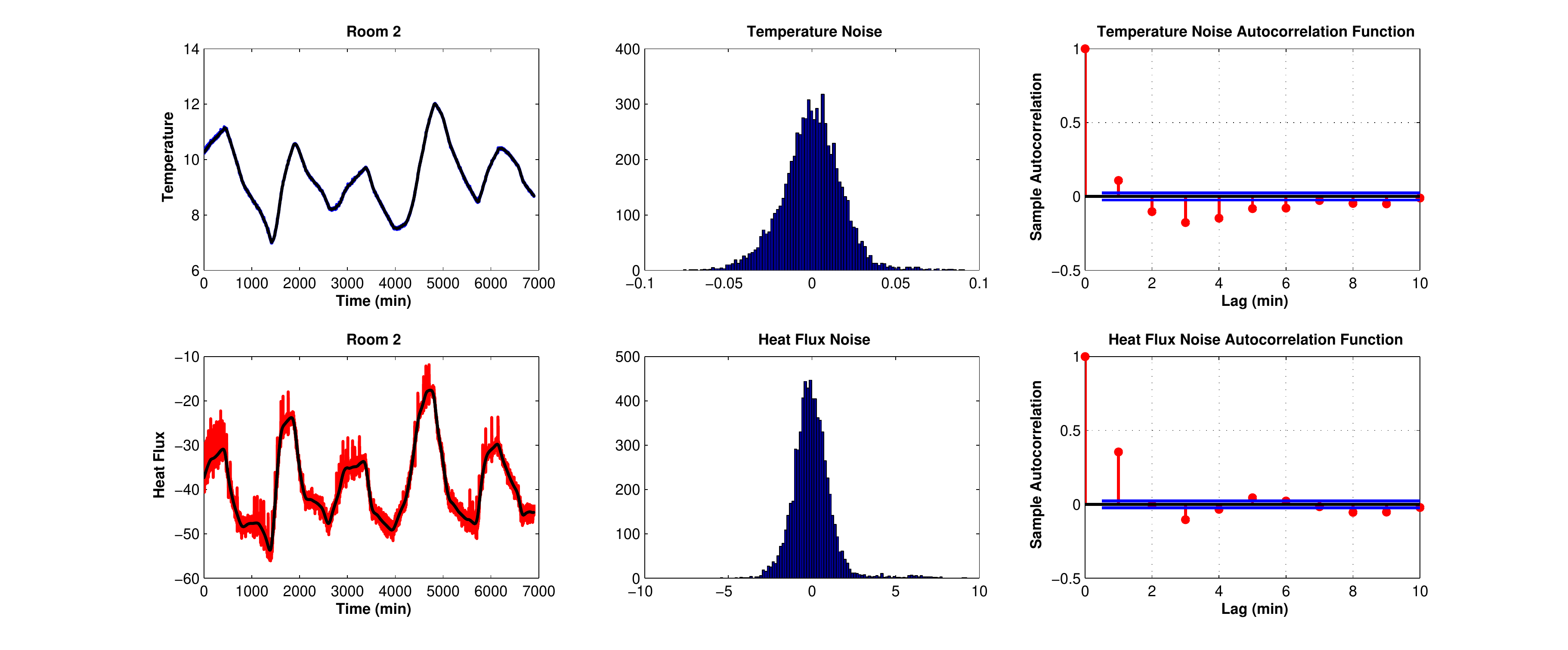}
\caption{Estimated noise of the raw temperature and heat flux measurements in Room 2.}
\label{fig4}
\end{figure}

\subsection{Smoothing spline method}
We assume that the time series $\mathbf{y} = (y_1, \ldots, y_N)$ follows the regression model
\begin{equation*}
y_i = g(t_i) + \epsilon_i , \,\,\,  i = 1, \ldots, N,
\end{equation*}
where $g(\cdot)$ is a smooth function that belongs to
\[W^{(m)}_{2} = \{ f: f^{(j)}\, \mbox{is absolutely continuous}, \, j = 0, 1, \ldots, m-1, \mbox{and}\, f^{(m)}\, \mbox{is square integrable} \}\]
and that $\epsilon_i$ are independent Gaussian random variables with zero mean and unknown variance, $\sigma^2$. We estimate $g$ by fitting a function to $\mathbf{y} = (y_1, \ldots, y_N)$ and adding a penalty measure of roughness:
\begin{equation*}
\min_{f \in W^{(m)}_{2}} \frac{1}{N} \sum_{i =1}^{N} (f(t_i) - y_i)^2 + \lambda \int ( f^{(m)}(u) )^2 du,
\end{equation*}
where $\lambda$ is the smoothing parameter. There are several methods on how to choose the smoothing parameter, $\lambda$ (see, for example, \cite{Wahba, hurvich}). Here, we consider a cubic smoothing spline estimator for $g$ where $m = 2$, which is computed by a MATLAB function (CSAPS). We choose $\lambda$ to minimize the autocorrelation function of the estimated noise.

\subsection{Data analysis}

\begin{figure}[h!]
\centering
\includegraphics[width=1.0\textwidth]{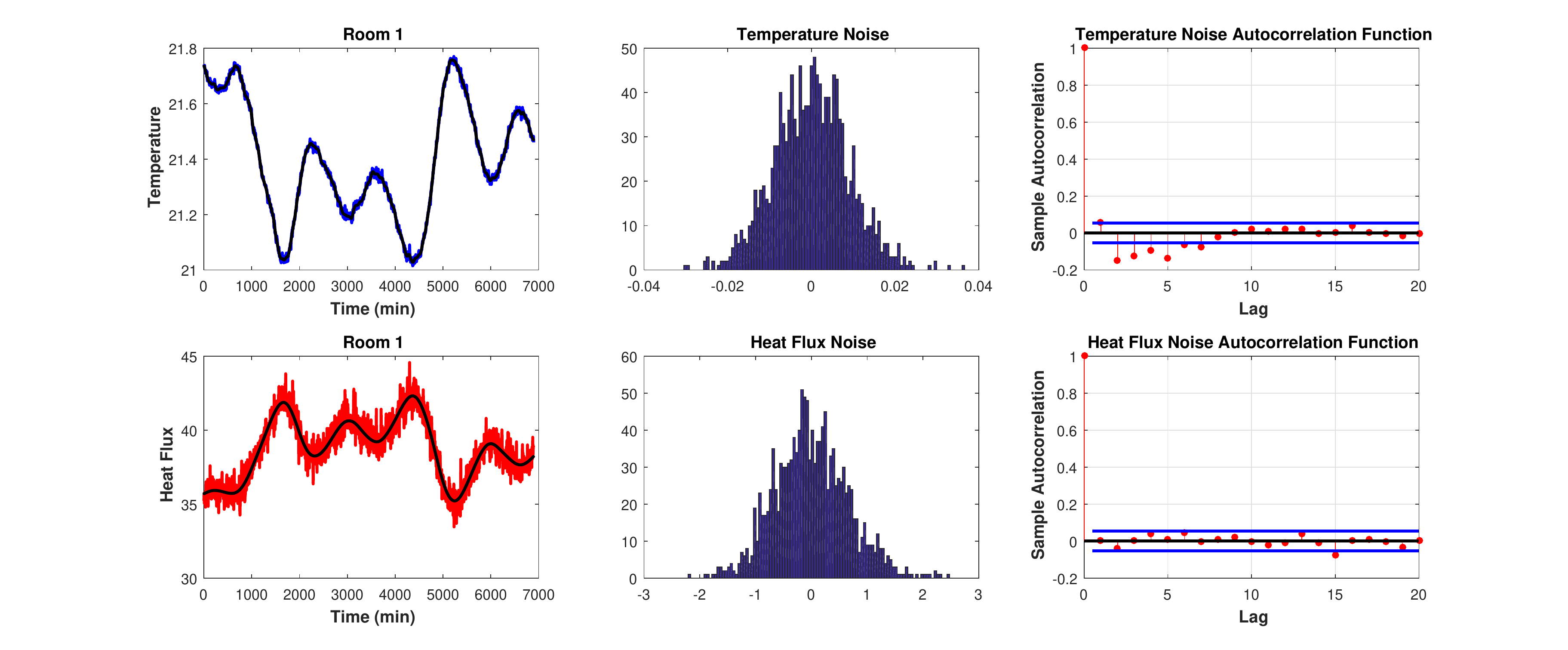}
\caption{Estimated noise of the moving average temperature and heat flux in Room 1.}
\label{fig5}
\end{figure}

\begin{figure}[h!]
\centering
\includegraphics[width=1.0\textwidth]{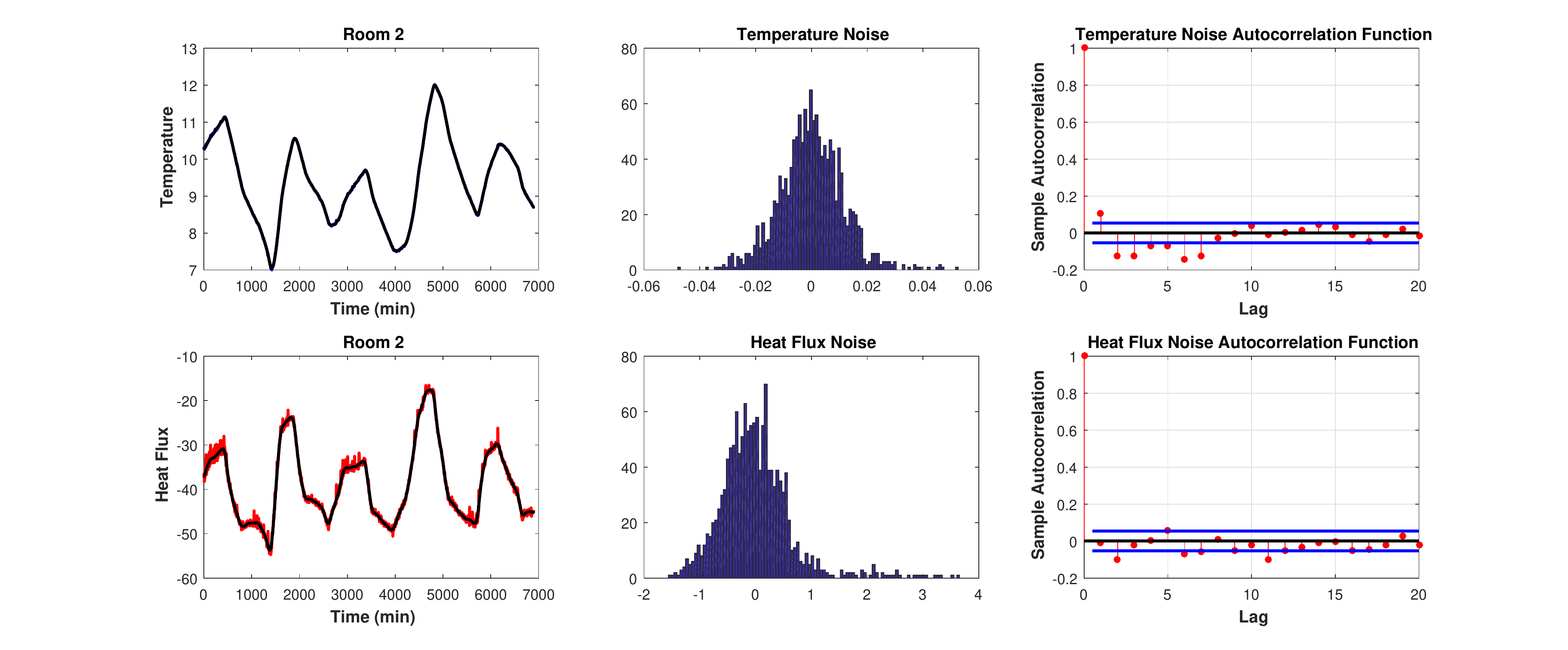}
\caption{Estimated noise of the moving average temperature and heat flux in Room 2.}
\label{fig6}
\end{figure}

Figures \ref{fig3} and \ref{fig4} show the estimated noise of the raw temperature and heat flux measurements on both sides of the wall. We notice that the estimated noise, especially in Room 1, is not Gaussian. Also, the autocorrelation function of the noise shows strong correlations, requiring the estimation of dense covariance matrices. We therefore consider a non-overlapping moving average of the raw data by computing local averages for every five consecutive measurements, where $5$ is the lag of the moving average. The lag 5 arises from the selection criterion that minimizes the total sum of the squared autocorrelation functions of the noise for the four time series. Figures \ref{fig5} and \ref{fig6} shows the estimated noise of the moving average temperature and heat flux, where we can see that the estimated noise looks Gaussian for all the time series. Moreover, the corresponding autocorrelations are considerably reduced. We replace the raw measurements shown in Figure \ref{fig1} with the moving average series and henceforth refer to these series as the data. Meanwhile, the original time series will be referred as raw data.

We distinguish between the moving average series based on their usage. The heat flux moving averages are used in the likelihood functions \eqref{JL} and \eqref{mar_li} associated with noise covariance matrices. Figures \ref{fig5} and \ref{fig6} show that the heat flux noises are almost stationary. This property can be tested by using the Ljung-Box Q-test on the two residual series (see \cite{box2015time}). Test results show evidence to accept the null hypothesis that the residual series are not autocorrelated. Therefore, we need only to estimate the corresponding variance of each residual series which is approximated by the sample variance with zero mean. On the other hand, the temperature moving averages are used to obtain data-driven priors \eqref{temp_Gauss}. In this case, we first approximate the mean functions, $\boldsymbol{\mu}_{int}$ and $\boldsymbol{\mu}_{ext}$ by means of smoothing splines. Then, we choose prior covariances that represent the uncertainties induced by the previous approximation. It is reasonable to assume that such uncertainties will be similar to the estimated noises of the temperature moving averages.  
 
 By replacing the original raw time series with the moving average series, we avoid estimating full covariance matrices of long time series which can be a difficult task, fulfilling the assumptions governing the statistical models \eqref{JL} and \eqref{temp_Gauss}. As a consequence, the number of observations is reduced from $6,900$ to $1,380$.

\subsection{Numerical Example 1}
\label{Ex1}
In this Subsection, we report a numerical example in which we use the Bayesian framework to infer the thermal properties of the wall by following a direct approach similar to the one in \cite{bidd} in which the boundary conditions, $\mathbf{T}_{int}$ and $\mathbf{T}_{ext}$, are assumed to be exact. However, in contrast to \cite{bidd} in which raw measurements are used, here $\mathbf{T}_{int}$ and $\mathbf{T}_{ext}$ are approximated by smoothing splines constructed from the boundary temperature measurements. The moving average heat fluxes, $\Q_{int}$ and $\Q_{ext}$, are assumed to be Gaussian distributed and uncorrelated. Moreover, $\Sigma_{int} = \sigma_{int}^{2} \mathcal{I}, \Sigma_{ext} =  \sigma_{ext}^{2} \mathcal{I}$, where $\sigma_{int} = \sigma_{ext} = 0.66$, which is the empirical standard deviation of the two moving average heat flux series. In this example, to get a likelihood function of $\theta$, we modify the joint likelihood \eqref{JL} by removing the dependency of $\mathbf{T}_{int}$ and $\mathbf{T}_{ext}$, which are assumed to be known. In other words, the likelihood is given by
\begin{equation}\label{like_1}
\mathcal{L}(\theta | \Q_{int}, \Q_{ext}) = \frac{1}{(2 \pi \sigma_{int} \sigma_{ext})^{N} } \exp \left\{ - \left( \frac{1}{2\sigma_{int}^{2}}||\Q_{int} - \F_{int}||_{2}^2 + \frac{1}{2\sigma_{ext}^{2}} ||\Q_{ext} - \F_{ext}||_{2}^2 \right) \right\} \,.
\end{equation}

Before applying Bayesian inference, we first compute the maximum likelihood estimate and validate the forward model with the raw measurements.

We maximize the likelihood function using a MATLAB function (FMINUNC) with several initial guess points. In the optimization algorithm, the heat equation is solved using $M=60$ in the space mesh for each time step, $\Delta t = 60$ seconds. The heat flux functions are computed by equations \eqref{fint}-\eqref{fext} for every five time steps. Then, the maximum likelihood (ML) estimates of the model parameters are
\[ R = 0.3107,\, \rho C = 3.17 \times 10^5,\, \tau_0 = 16.11 \, . \]

\begin{figure}[h!]
\centering
\begin{minipage}{.45\textwidth}
  \centering
  \includegraphics[width=9cm]{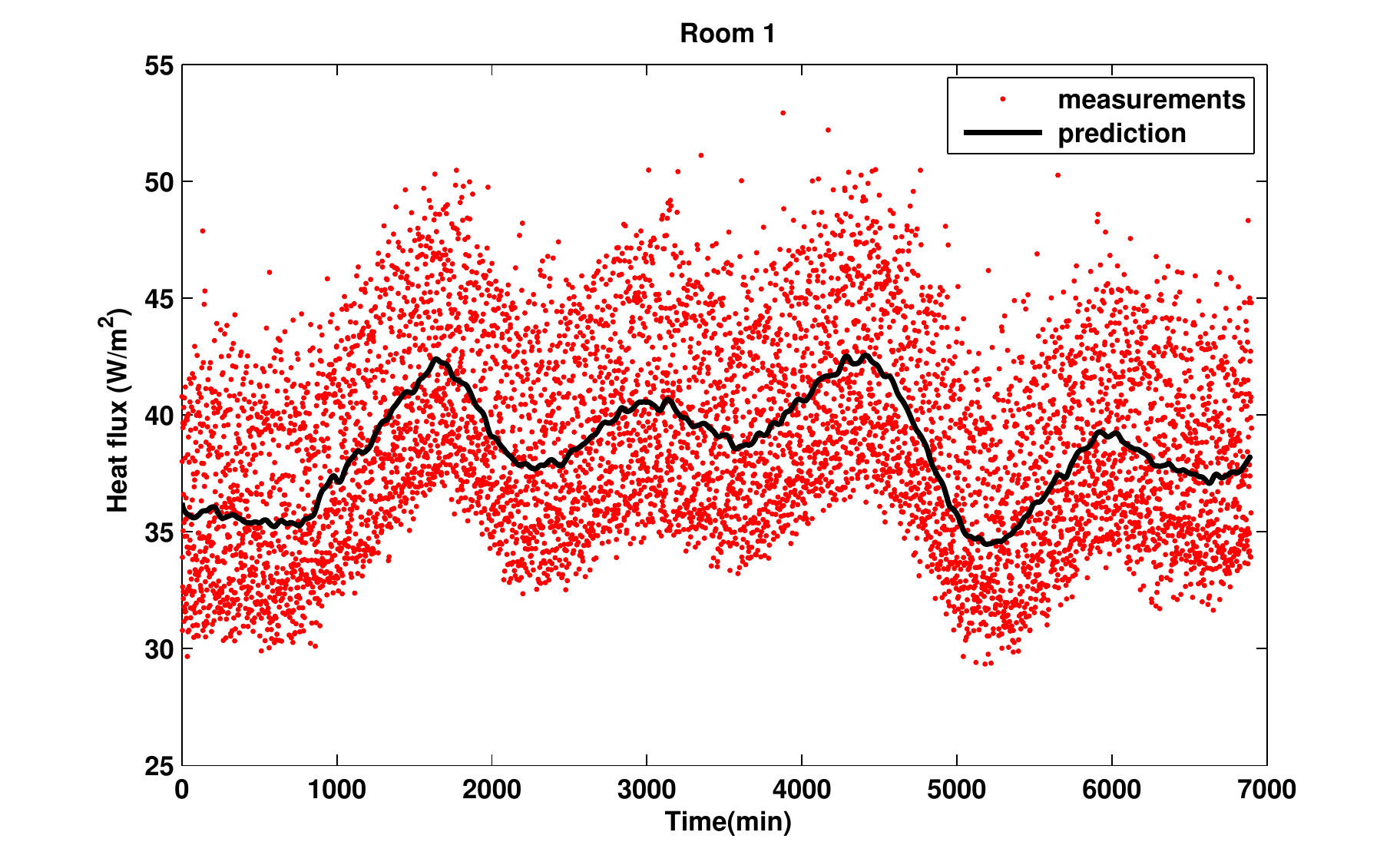}
  \captionof{figure}{Raw heat flux measurements (red dots) with a model prediction for Room 1.}
  \label{pred1}
\end{minipage}
~
\begin{minipage}{.45\textwidth}
  \centering
  \includegraphics[width=9cm]{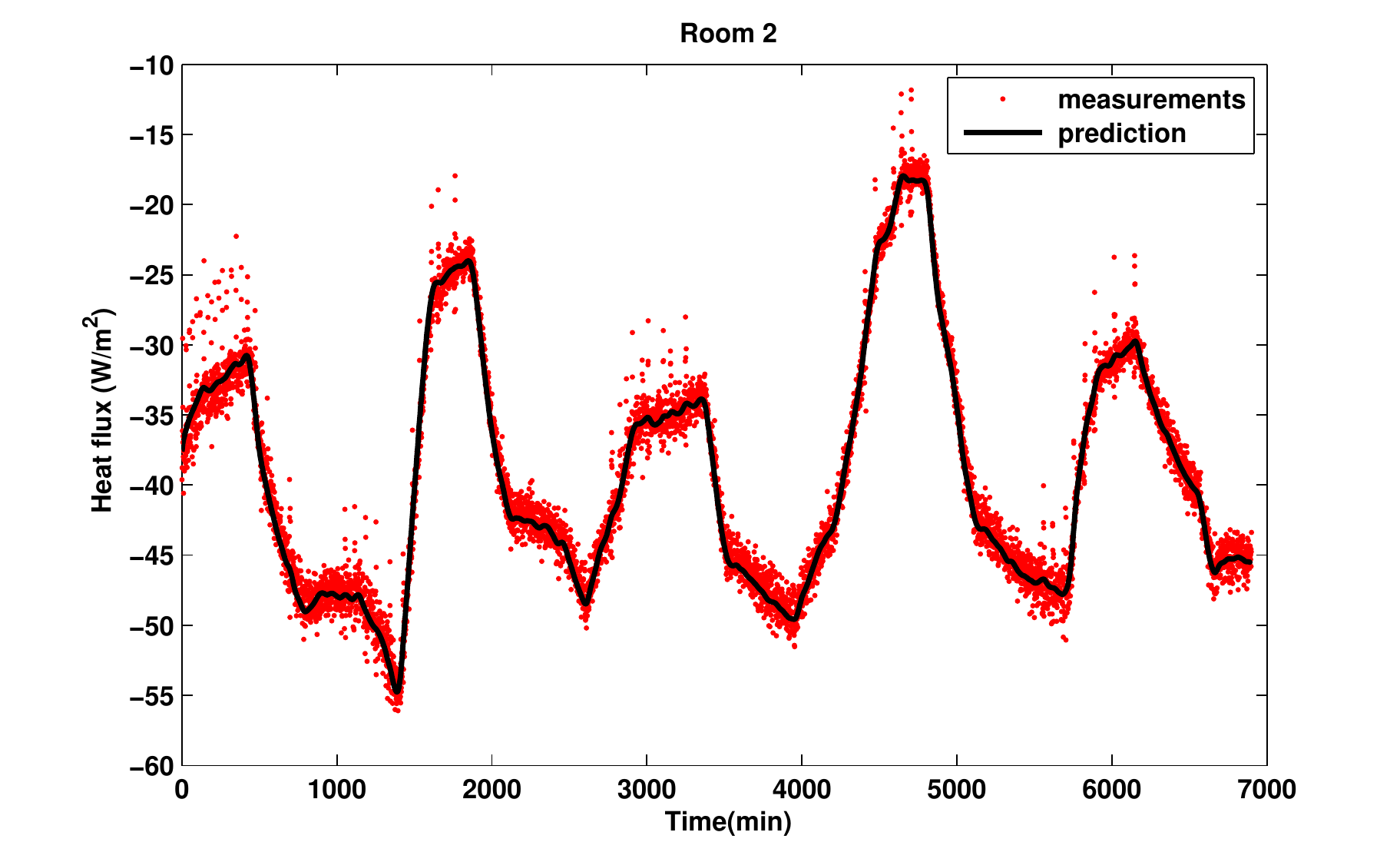}
  \captionof{figure}{Raw heat flux raw measurements (red dots) with a model prediction for Room 2.}
   \label{pred2}
\end{minipage}
\end{figure}

Note that the values of $R$ and $\rho C$ are well within the range provided by the tabulated values as described at the beginning of this section. We also assess the consistency of such ML estimates by plugging them into the forward model to compare the simulated heat flux with the experimental measurements. Figures \ref{pred1} and \ref{pred2} show that the heat flux simulations, computed by using the above ML estimates of $\theta$, are consistent with the data.

We now take the Bayesian approach by computing the posterior distribution of $\theta$ associated with the likelihood (\ref{like_1}):
\begin{equation*}
\pi(\theta | \Q_{int}, \Q_{ext}) \propto \mathcal{L}(\theta | \Q_{int}, \Q_{ext}) \pi_{p}(\theta)\, .
\end{equation*}
To specify the joint prior, $\pi_{p}(\theta)=\pi_{p}(R,\rho C , \tau_{0})$, we assume independence among the parameters and we consider the following uniform priors:
\[R \sim U(0.17,0.36), \rho C  \sim U(234000, 431000), \tau_0 \sim U(5, 25)\,.\]

The marginal posterior densities of $R, \rho C$ and $\tau_0$ are obtained by using the Laplace method and a Markov chain Monte Carlo (MCMC) sampling algorithm. The Laplace method provides a Gaussian approximation of the posterior distribution of $\theta$ as follows:
\begin{equation*}
\pi(\theta| \mathbf{Q}_{int}, \mathbf{Q}_{ext}) \approx \frac{1}{\sqrt{(2 \pi)^{3} |H(\hat{\theta})|}} \exp \left\{ -(\theta- \hat{\theta})^{tr}H(\hat{\theta})^{-1}(\theta- \hat{\theta}) \right\} \, ,
\end{equation*}
where $\hat{\theta}$ is the maximum a posteriori (MAP) probability estimate of $\theta$ and $H(\hat{\theta})$ is the inverse Hessian matrix of the negative log posterior evaluated at $\hat{\theta}$ \cite[Chapter 4]{Ghosh}. 

\begin{figure}[h!]
\centering
\includegraphics[width=1.0\textwidth]{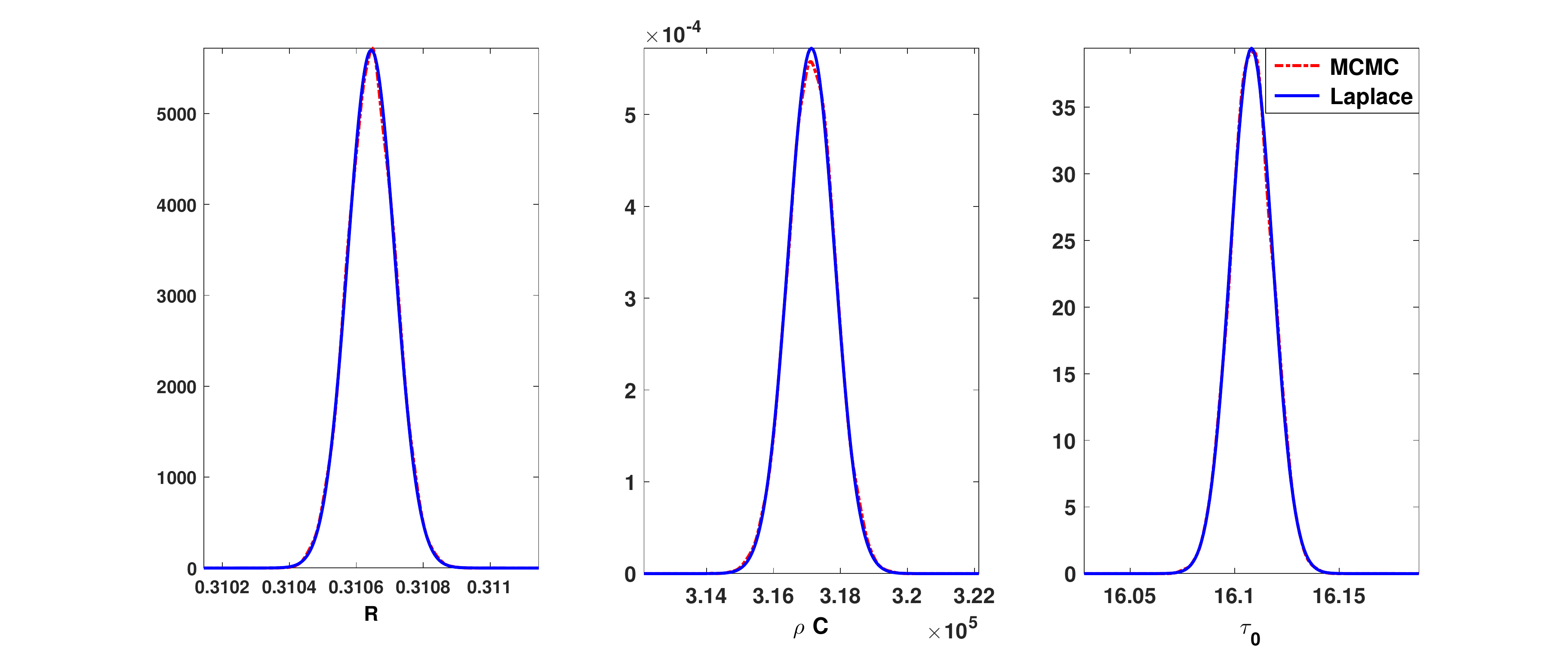}
\caption{Marginal posteriors of the model parameters $R, \rho C$ and $\tau_0$ approximated by the Laplace method (blue line) and the random-walk Metropolis-Hastings algorithm (red line).}
\label{laplace1}
\end{figure}

We used the random-walk Metropolis-Hastings algorithm (\ref{alg1}) to generate MCMC samples (see \cite[Chapter 6]{robert} and \cite{babuvska2016bayesian}). We ran the MCMC chain $101,000$ times, with a $1,000-$iteration burn-in period and every twentieth draw of the chain kept. Figure \ref{laplace1} shows that the Laplace method provides a very accurate approximation of the three posterior densities when compared with the simulation-based posterior densities. The marginal posterior densities of $R$ and $\tau_0$ are highly concentrated around their respective modes. 

\begin{algorithm}[h!]
\caption{Random-walk Metropolis-Hastings algorithm} \label{alg1}
\begin{algorithmic}[1]
\State \textbf{set} an initial value for the chain: $\theta_c = \theta_0$ and \textbf{choose} the covariance $diag(\delta)$
\State \textbf{run} the forward model at $\theta_c$ up to time $t_N$
\State \textbf{compute} $a = loglikelihood(\theta_c) + logprior(\theta_c)$
\State \textbf{draw} $\theta_p$ from $N(\theta_c, diag(\delta))$
\State \textbf{run} the forward model at $\theta_p$ up to time $t_N$
\State \textbf{compute} $b = loglikelihood(\theta_p) + logprior(\theta_p)$
\State \textbf{let} $ H = min(1,exp(b-a))$ and \textbf{draw} $r$ from $U(0,1)$
\If{$H > r$}
\State $\theta_c = \theta_p$
\State $ a = b $
\EndIf
\State \textbf{repeat} steps (2 to 10) \textbf{until} $S$ posterior samples are obtained.
\end{algorithmic}
\end{algorithm}

\subsection{Numerical Example 2}
\label{Ex2}
In this example, we incorporate uncertainty in the observations of $\T_{int}$ and $\T_{ext}$ and we apply our proposed hierarchical approach to characterize the posterior distribution of $\theta$ that arises from the marginalized likelihood \eqref{mar_li}. More precisely, we assume that the nuisance boundary conditions, $\T_{int}$ and $\T_{ext}$, are modeled by the Gaussian distributions introduced in \eqref{temp_Gauss}, where $\boldsymbol{\mu}_{int}$ and $\boldsymbol{\mu}_{ext}$ are the smoothing splines constructed from the boundary temperature data and we let $C_{int,p} = C_{ext,p} = (0.01)^{2} \mathcal{I}$ using the estimated noises of the moving average temperature series. The initial condition is approximated by the piecewise linear function \eqref{model_as} using the initial surface temperature measurements. Similar to Example \ref{Ex1}, the moving averages, $\Q_{int}$ and $\Q_{ext}$, for heat flux are assumed to be Gaussian distributed and uncorrelated with $\Sigma_{int} = \Sigma_{ext} = (0.66)^{2} \mathcal{I}$.

\begin{figure}[h!]
\centering
\begin{minipage}{.45\textwidth}
  \centering
  \includegraphics[width=9cm]{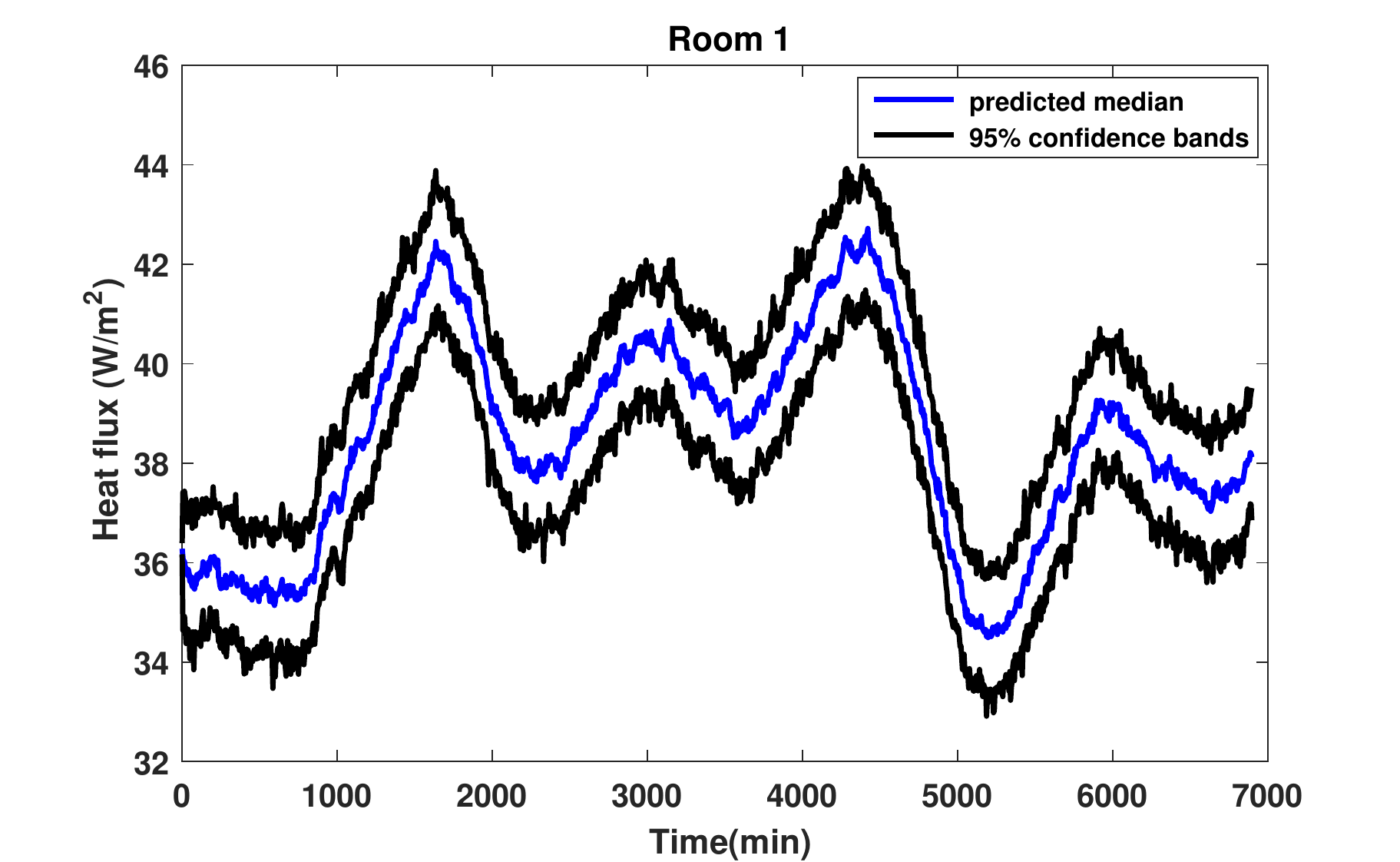}
  \captionof{figure}{Median prediction (blue line) and $95 \%$ confidence bands (black lines) for the heat flux in Room 1.}
  \label{pred3}
\end{minipage}
~
\begin{minipage}{.45\textwidth}
  \centering
  \includegraphics[width=9cm]{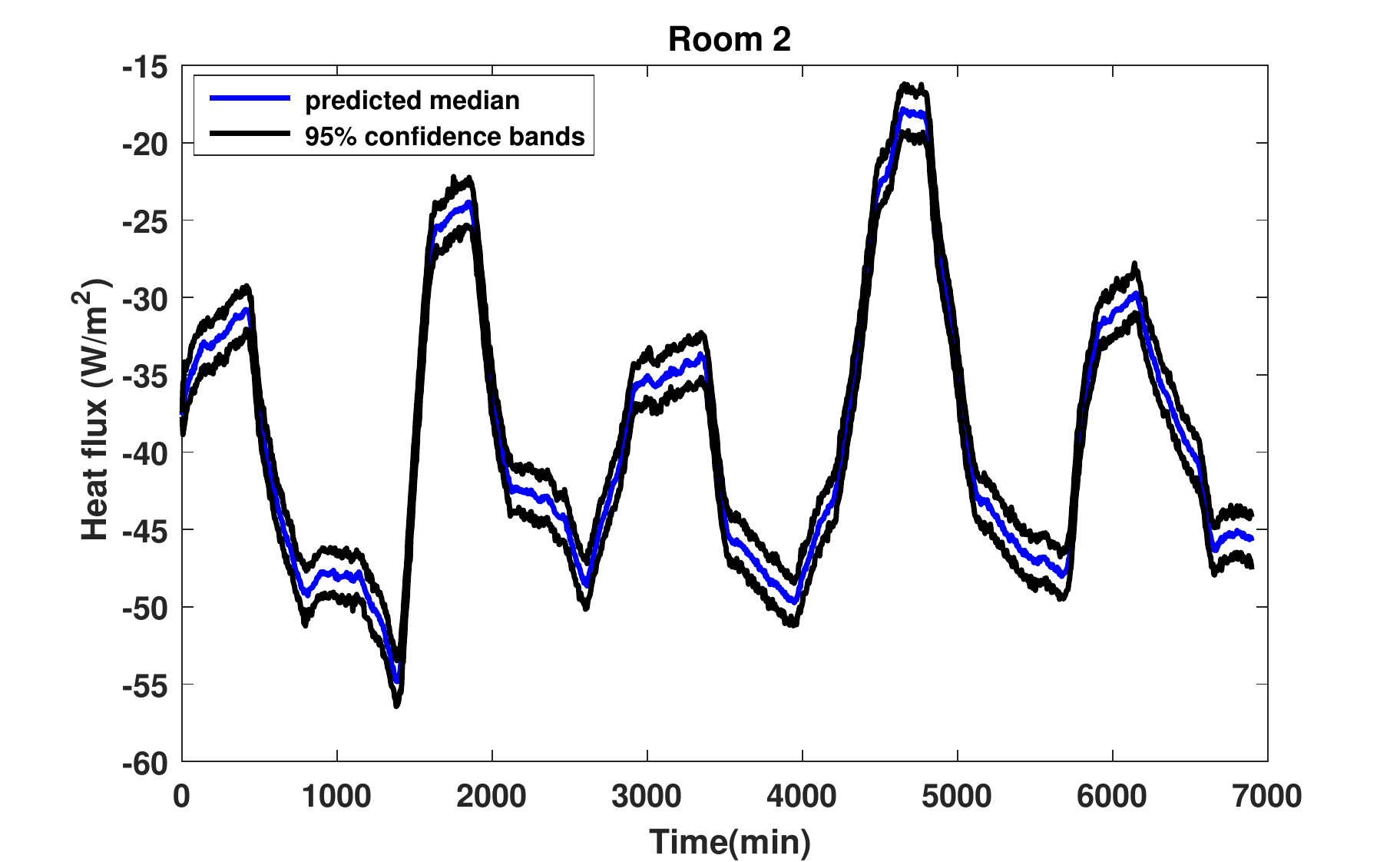}
  \captionof{figure}{Median prediction (blue line) and $95 \%$ confidence bands (black lines) for the heat flux in Room 2.}
   \label{pred4}
\end{minipage}
\end{figure}

The ML estimates of the components of $\theta$ corresponding to the marginal likelihood \eqref{mar_li} are
\[ R = 0.3106,\, \rho C = 3.20 \times 10^5, \, \tau_0 = 16.11\,. \]
Given the ML estimates, we plot the predicted median heat flux with $95 \%$ confidence bands in Figures \ref{pred3} and \ref{pred4}, where the boundary conditions are sampled from
\[ \T_{int} \sim N( \boldsymbol{\mu}_{int},  (0.01)^{2} \mathcal{I} ),\qquad  \T_{ext} \sim N( \boldsymbol{\mu}_{ext},  (0.01)^{2} \mathcal{I} )\, .\]

\begin{figure}[h!]
\centering
\includegraphics[width=1.0\textwidth]{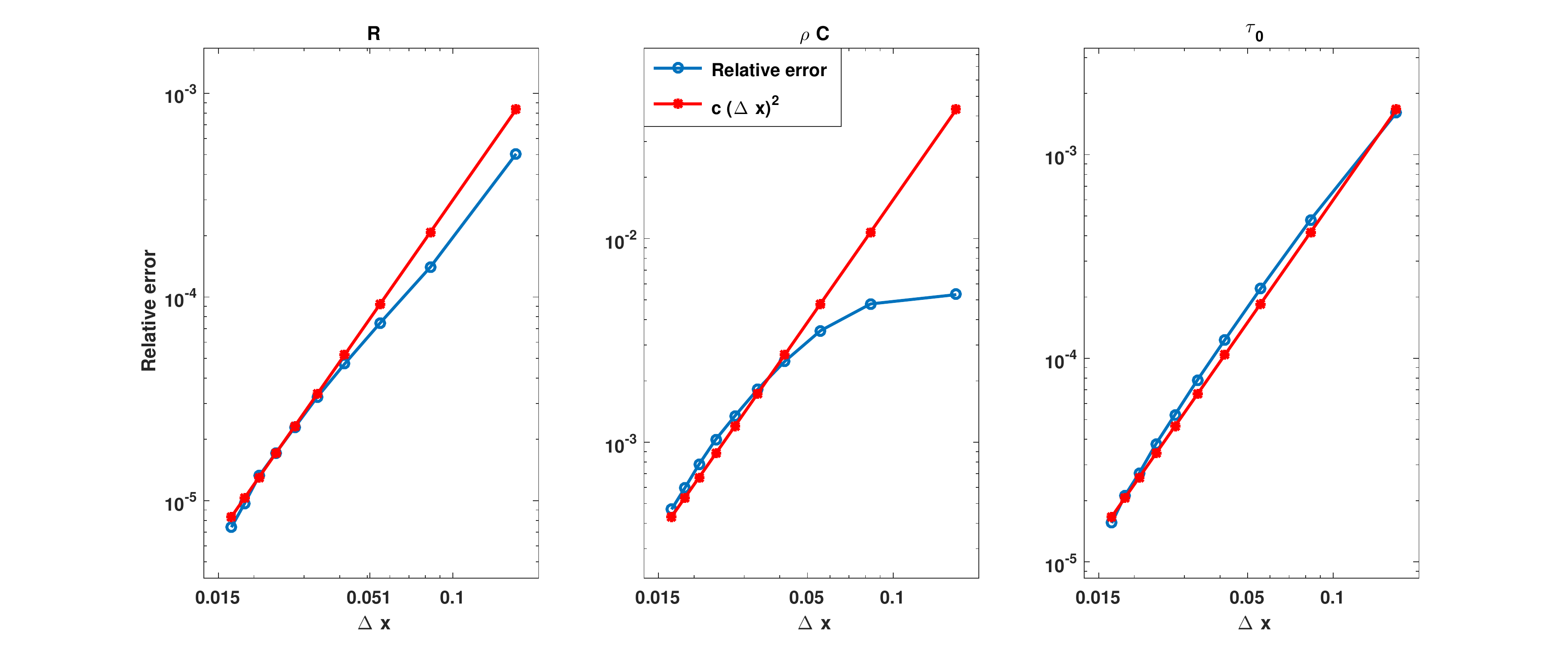}
\caption{Convergence analysis of ML estimates with respect to $\Delta x$.}
\label{num_con}
\end{figure}

\begin{figure}[h!]
\centering
\includegraphics[width=1.0\textwidth]{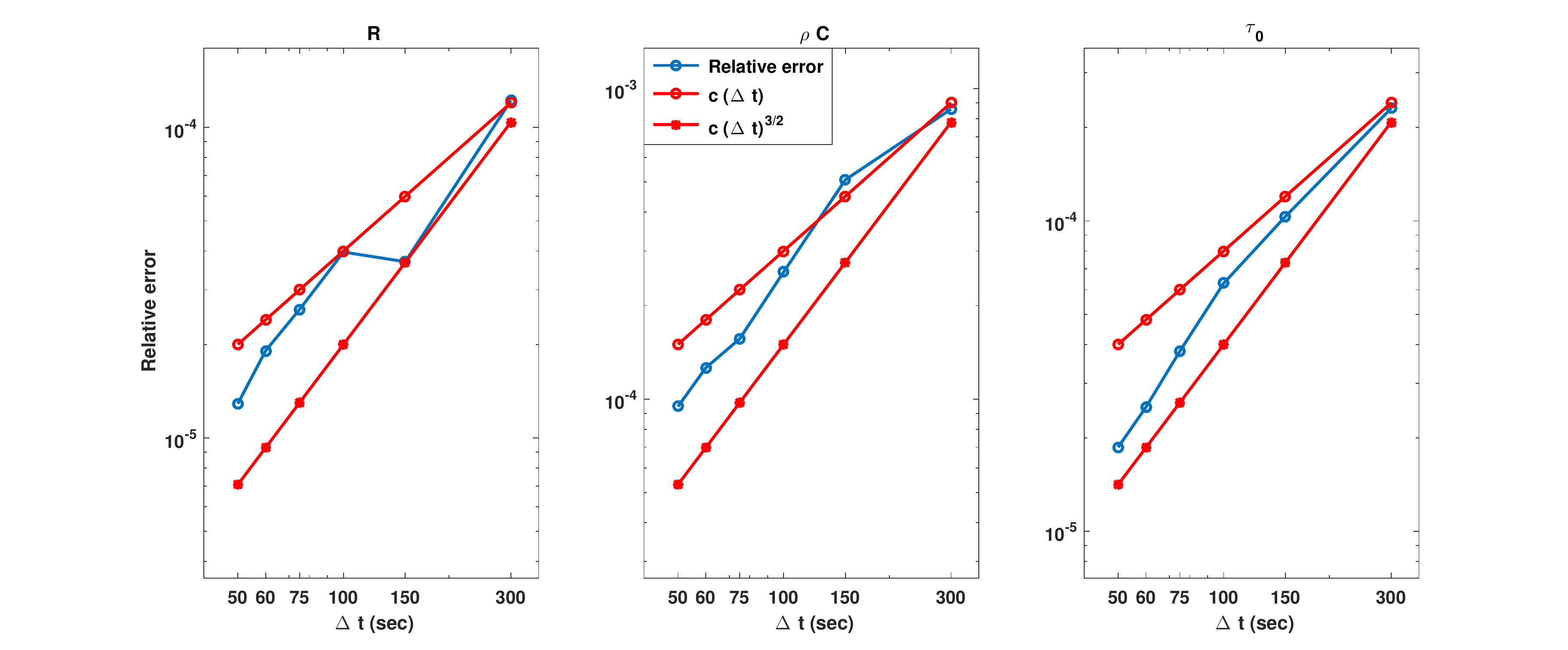}
\caption{Convergence analysis of ML estimates with respect to $\Delta t$.}
\label{num_cont}
\end{figure}

One of the main contributions of our work, relative to existing Bayesian approaches that infer the thermal properties of walls, is that we use the heat equation to describe heat transfer through the wall. Analysis of the effect of the space-time discretization $(\Delta x, \Delta t)$ on ML estimates of the components of $\theta$ is, therefore, necessary. We determined that the convergence of the ML estimates is quadratic with respect to $\Delta x$ and linear with respect to $\Delta t$ as shown in Figures \ref{num_con} and \ref{num_cont}, respectively. We also study the effect of the initial condition approximation by estimating the thermal properties and comparing the corresponding Akaike information criterion (AIC) for different initial conditions. The smallest AIC value indicates the preferred model among possible models \cite{aicc}. Table \ref{AIC} shows that the piecewise linear approximation of the initial condition is better than linear and higher order approximations.

\begin{table}[h!]
\centering
\begin{tabular}{|c|c|c|c|}
\hline
Initial condition &  $R$  &  $\rho C \times 10^5$ &  AIC  \\
\hline
Linear &  0.3106  &  3.197 &  -25020\\
\hline
Piecewise linear & 0.3106  &  3.20    &  -25066 \\
\hline
Quadratic & 0.3106  &  3.20  &  -25056 \\
\hline
Cubic & 0.3105  &  3.196  &  -24973  \\
\hline
\end{tabular} 
\captionof{table}{ML estimates of thermal resistance and heat capacity under different initial conditions.}
\label{AIC}
\end{table}

\begin{figure}[h!]
\centering
\includegraphics[width=1.0\textwidth]{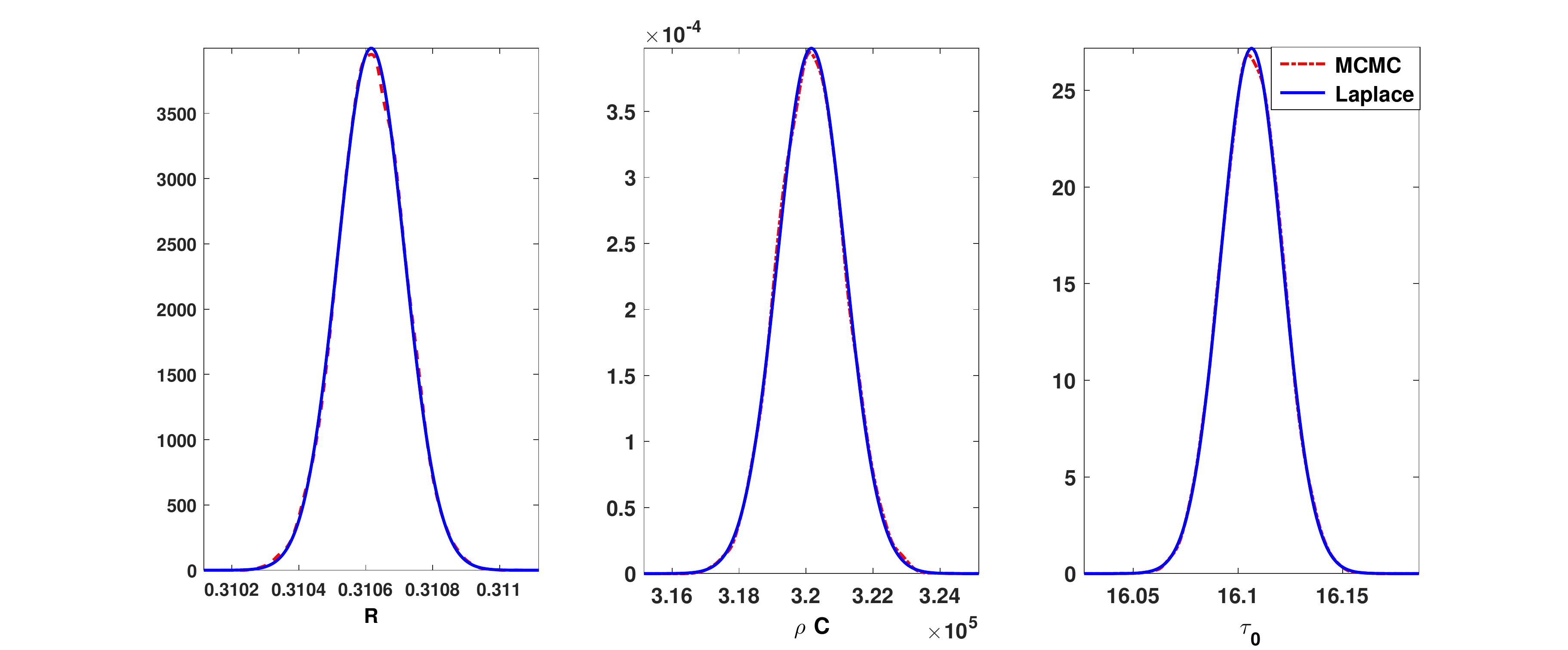}
\caption{Marginal posteriors of $R, \rho C$ and $\tau_0$ approximated by the Laplace method (blue line) and the random-walk Metropolis-Hastings algorithm (red line).}
\label{laplace2}
\end{figure}

\begin{figure}[h!]
\centering
\begin{minipage}{.45\textwidth}
  \centering
  \includegraphics[width=9cm]{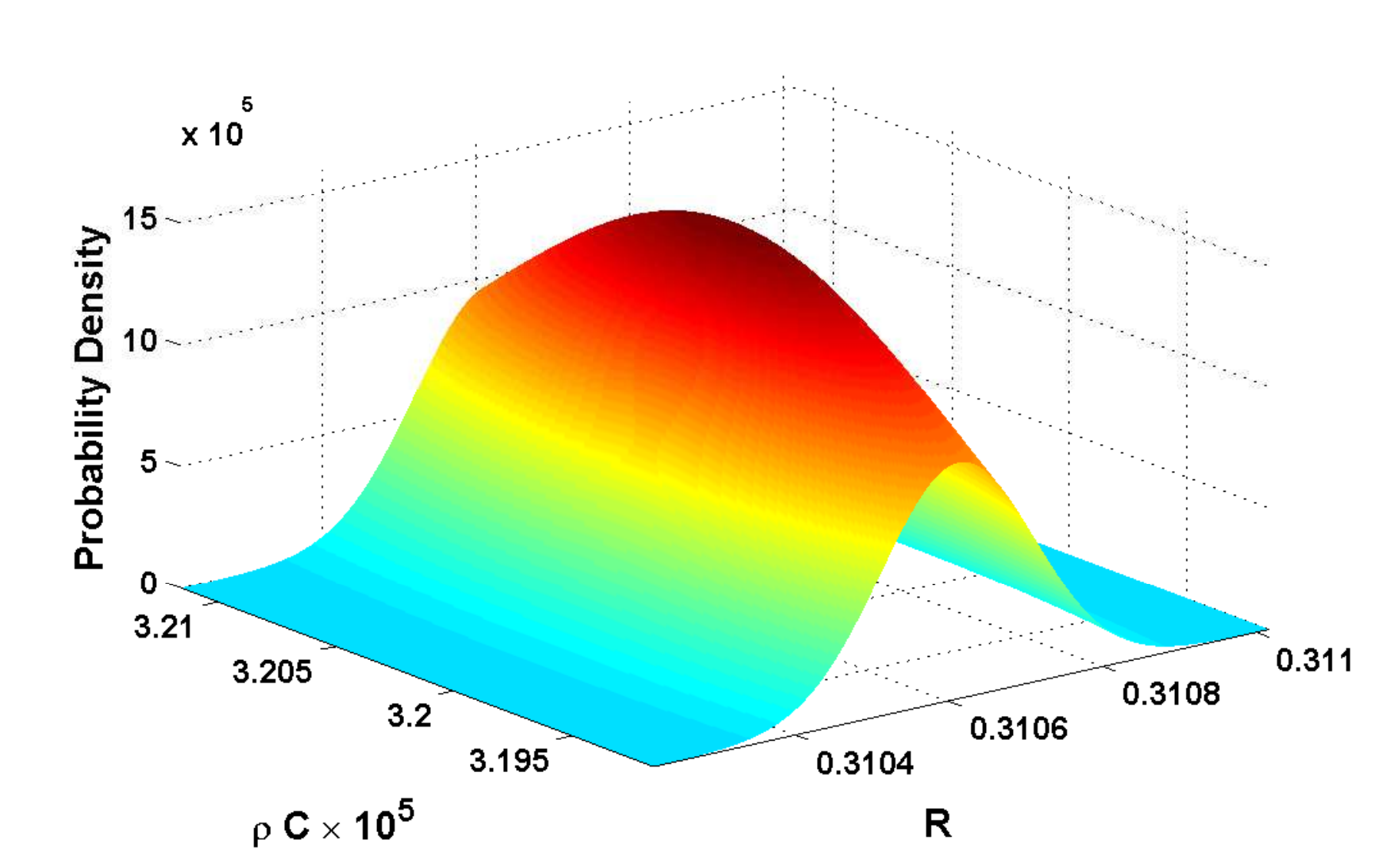}
  \captionof{figure}{Approximated bivariate posterior distribution of $R$ and $\rho C$.}
  \label{biv2}
\end{minipage}
~
\begin{minipage}{.45\textwidth}
  \centering
  \includegraphics[width=9cm]{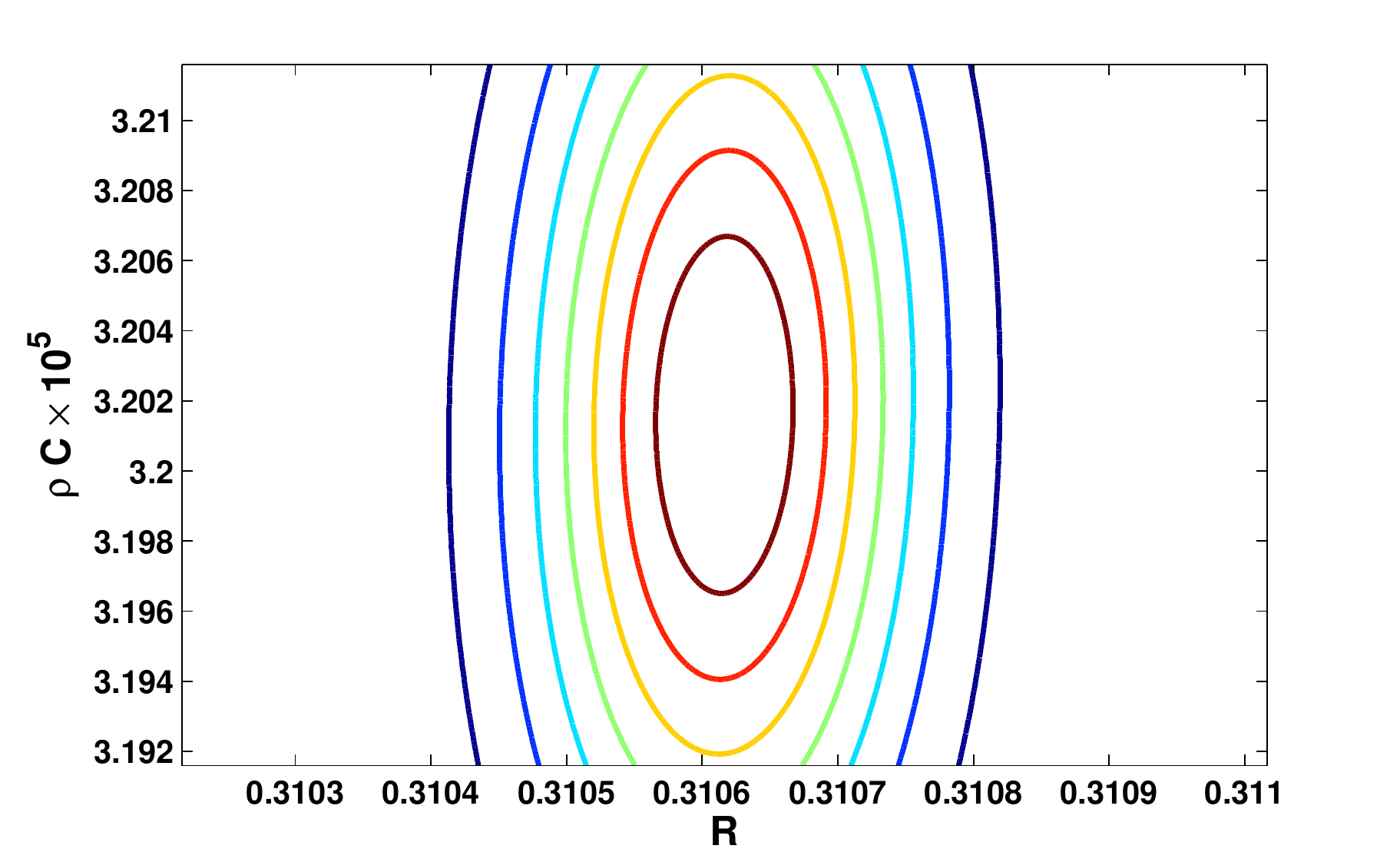}
  \captionof{figure}{Contour plot of the approximated bivariate posterior distribution of $R$ and $\rho C$.}
   \label{cont2}
\end{minipage}
\end{figure}

\begin{figure}[h!]
\centering
\includegraphics[width=1.0\textwidth]{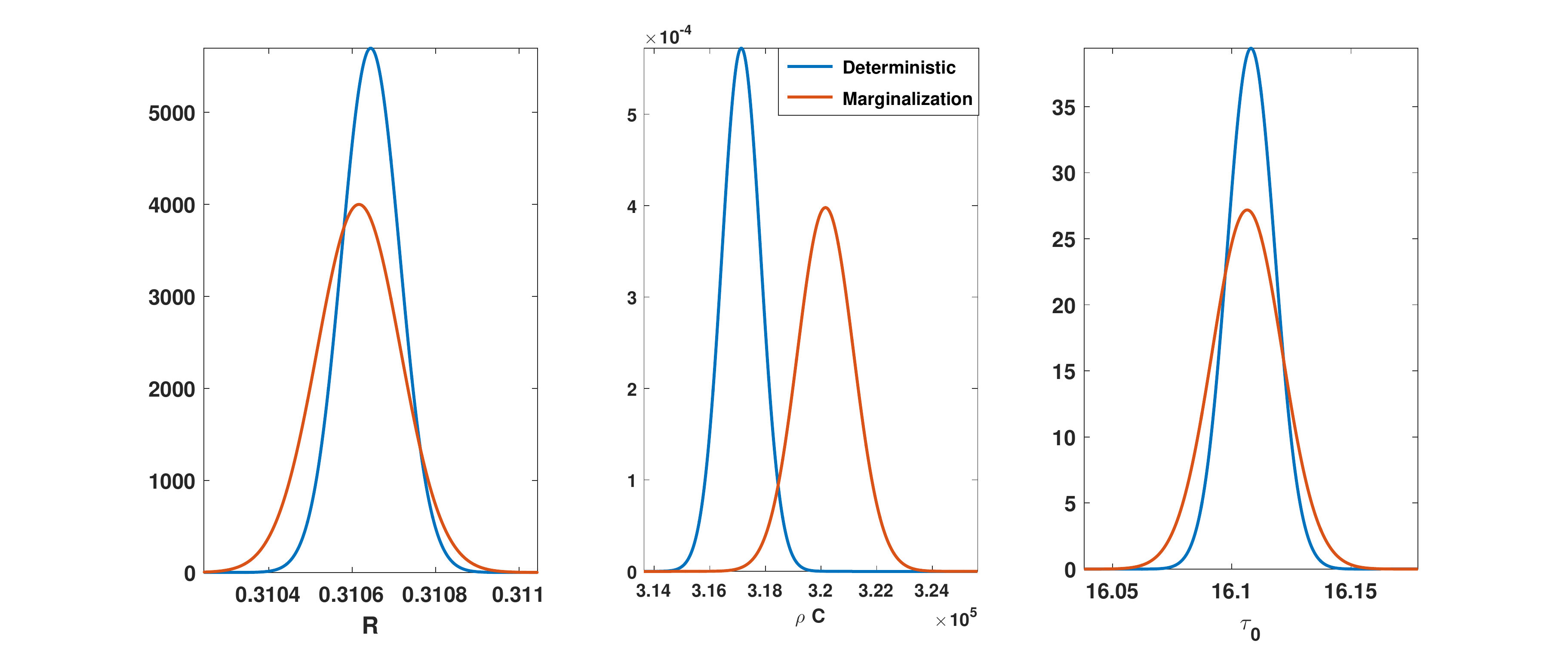}
\caption{Comparison between the marginal posteriors obtained by the deterministic approach (Figure \ref{laplace1}) and the marginalization approach (Figure \ref{laplace2}).}
\label{laplace3}
\end{figure}

We now consider the Bayesian approach using the marginal likelihood defined in (\ref{mar_li}) with the following uniform priors: 
\[ R \sim U(0.17,0.36), \rho C  \sim U(234000, 431000), \tau_0 \sim U(5, 25)\,.\]
We again obtain the corresponding marginal posterior densities by the Markov chain Monte Carlo (MCMC) sampling algorithm (Algorithm \ref{alg1}) and by using the Laplace method. Figure \ref{laplace2} shows the estimated marginal posteriors for $R$, $\rho c$ and $\tau_0$. The Laplace method and the MCMC technique provide very similar estimated marginal posterior densities. Also, Figures \ref{biv2} and \ref{cont2} show the correlation between the thermal resistance, $R$, and the heat capacity, $\rho C$. In Figure \ref{laplace3}, we compare these marginal posteriors with the ones obtained in Subsection \ref{Ex1} where the boundary parameters are assumed to be deterministic. The direct deterministic approach provides over-concentrated posteriors and relatively biased MAP estimates. On the other hand, the marginalization approach incorporates uncertainties into the nuisance boundary parameters, thereby producing realistic posterior densities. 

\begin{figure}[h!]
\centering
\includegraphics[width=1.0\textwidth]{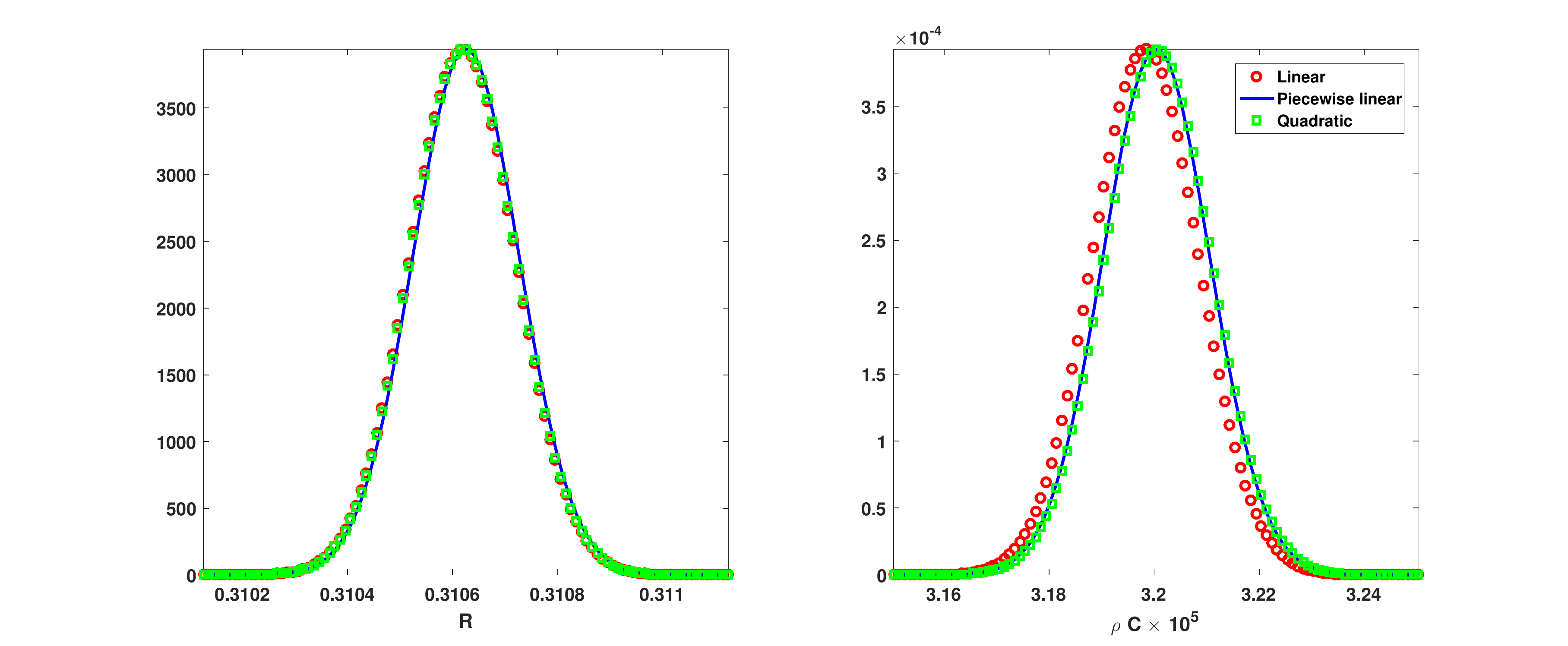}
\caption{Comparison between the marginal posteriors of $R$ and $\rho C$ obtained by the marginalization approach under different approximations of the initial condition.}
\label{Initialcond}
\end{figure}

We also analyze the effect of the initial condition approximation by comparing the marginal posteriors of the thermal properties. Figure \ref{Initialcond} shows that the posteriors of $R$ and $\rho C$ are very similar under three different initial conditions.

\subsection{Robustness analysis}
\label{Robust}

To assess the robustness of our Bayesian approach, we consider a subsampling method that generates variability intervals for $R$ and $\rho C$. First, the raw time series is divided into consecutive, non-overlapping subintervals of length $\ell$. Then, we resample the original time series in which subsamples of size $b$ are drawn from each subinterval. The local average is computed for each subsample to filter the sampled time series. We use smoothing splines to model the boundary temperature parameters. We then use our Bayesian approach under the same uniform priors used in Examples \ref{Ex1} and \ref{Ex2}. By repeating this procedure, we obtain several MAP estimates. We summarize the variability of these estimates by means of boxplots. The subsampling procedure is summarized in Algorithm \ref{alg2}.

\begin{algorithm}[h!]
\caption{Subsampling algorithm} \label{alg2}
\begin{algorithmic}[1]
\State \textbf{partition} the raw time series into subintervals $D_i$, $i = 1, \ldots, \ell$
\State \textbf{sample} b observations from each $D_i$ without replacement
\State \textbf{compute} local averages for each subsample
\State \textbf{estimate} $\boldsymbol{\mu}_{int}$ and $\boldsymbol{\mu}_{ext}$ using smoothing spline fit of the averaged time series
\State \textbf{apply} the Bayesian inference given the averaged heat flux measurements
\State \textbf{repeat} steps (2 to 4) \textbf{until} $S$ MAP estimates are obtained.
\end{algorithmic}
\end{algorithm}

\begin{figure}[h!]
\centering
\includegraphics[width=0.95\textwidth]{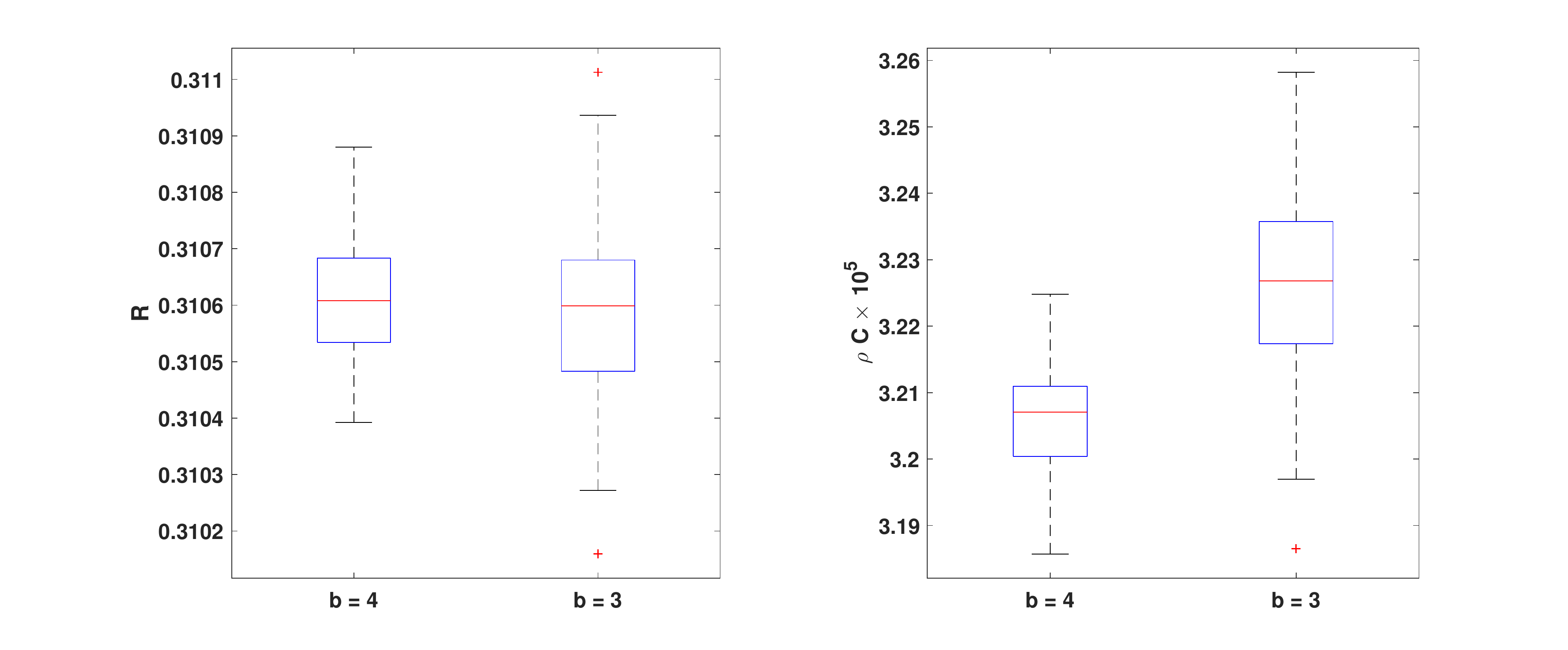}
\caption{The variability of $R$ and $\rho C$ using subsampling algorithms with subinterval length $\ell = 5$ and subsample sizes $b = 4$ and $3$.}
\label{boxplot}
\end{figure}

\begin{figure}[h!]
\centering
\includegraphics[width=0.95\textwidth]{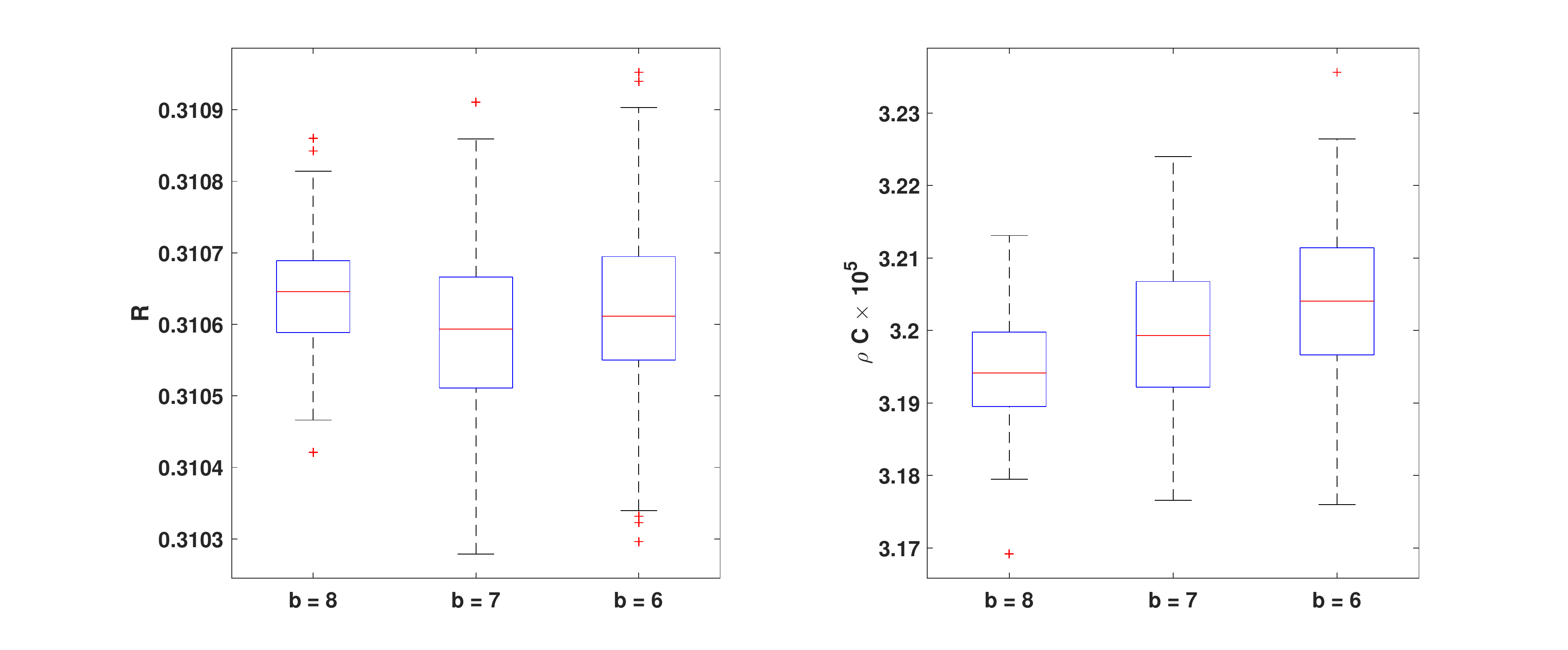}
\caption{The variability of $R$ and $\rho C$ using subsampling algorithms with subinterval length $\ell = 10$ and subsample sizes $b = 8, 7$ and $6$.}
\label{boxplot2}
\end{figure}

Figures \ref{boxplot} and \ref{boxplot2} show the variability intervals obtained for $R$ and $\rho C$ using the subsampling algorithm that draws $b$ observations randomly from each subinterval of size $\ell$. In Figure \ref{boxplot}, we use $\ell = 5$ and compare the variability between drawing $4$ and $3$ observations. Clearly, the uncertainty increases when small subsamples are used. Similarly, Figure \ref{boxplot2} shows the difference in variability between sampling algorithms that randomly draw $8, 7$ and $6$ observations from each subinterval. In general, the variability intervals include our MAP estimate of $\theta$ obtained in Example \ref{Ex2}. These intervals are within a reasonable range. Such results ensure the robustness of our Bayesian methodology in estimating the thermal properties of a wall.

\section{Information gain}
\label{InfGain}

In a Bayesian framework, the utility of an experiment can be measured by the so-called information gain (relative entropy) function, which is defined by the Kullback-Leibler divergence between the prior density function, $\pi_{p}(\theta)$, and the posterior density function, $\pi(\theta|\mathbf{Q}_{int}, \mathbf{Q}_{ext}, \xi)$:

\begin{equation}
D_{KL}(\mathbf{Q}_{int}, \mathbf{Q}_{ext}, \xi) := \int_{\Theta} \log \left( \frac{\pi(\theta|\mathbf{Q}_{int}, \mathbf{Q}_{ext}, \xi)}{\pi_{p}(\theta)} \right) \pi(\theta|\mathbf{Y}, \xi) d\theta \,, \label{eq:kulldiv}
\end{equation}

where $\xi$ is the experimental setup \cite{Kull, Quan}.

\begin{figure}[h!]
\centering
\includegraphics[width=0.7\textwidth]{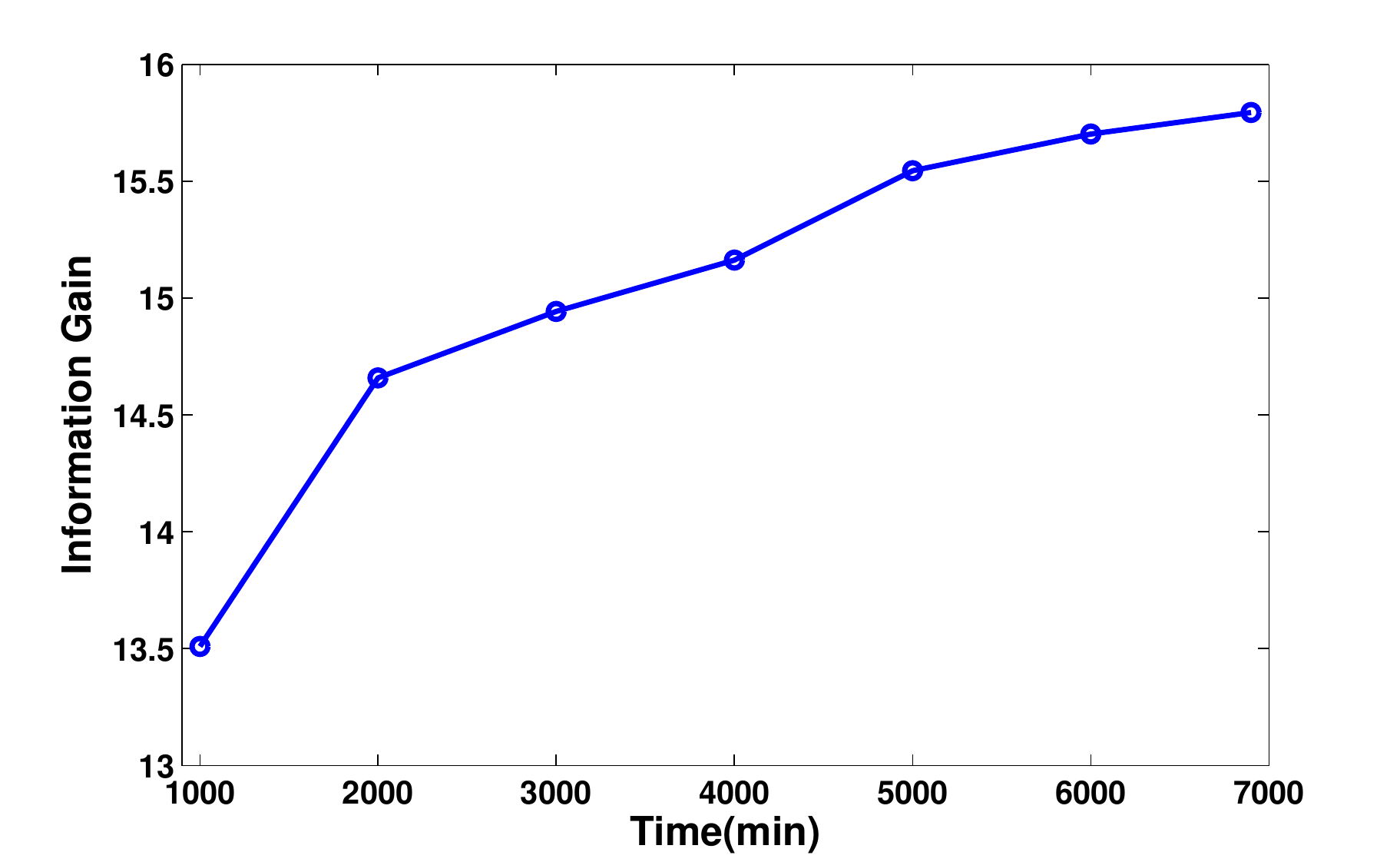}
\caption{The estimated information gain with respect to time.}
\label{inf_gain}
\end{figure}

We first consider an experimental setup that describes the duration of the measurement campaign. We use the Laplace approximation to compute the information gain for overlapping increasing time intervals. Figure \ref{inf_gain} shows that the information gain increases over time as more observations are incorporated into the Bayesian inference. However, we observe that after $5000$ minutes, the rate of increase of the information gain slows, indicating that the collected measurements provide reliable information on the thermal properties of the solid brick wall. We introduce another experimental setup by considering the external temperature oscillation. Figure \ref{inf_gain_cycles} shows how data are partitioned on the basis of the detected external temperature cycles in Room 2. We estimate the information gain for each cycle, and the results are summarized in Table \ref{infgain_cycles}. Cycles 1 and 3 are ranked as the most informative in terms of the Kullback-Leibler divergence, while Cycle 2 is the least informative, although Cycle 4 has the shortest duration. From these results, we can deduce that larger temperature oscillations bring more knowledge to the inference about $\theta$, suggesting that steady state conditions cannot be used to infer the thermal properties of the wall.

\begin{minipage}{\textwidth}
\begin{minipage}[b]{0.49\textwidth}
\centering
\includegraphics[width=8cm]{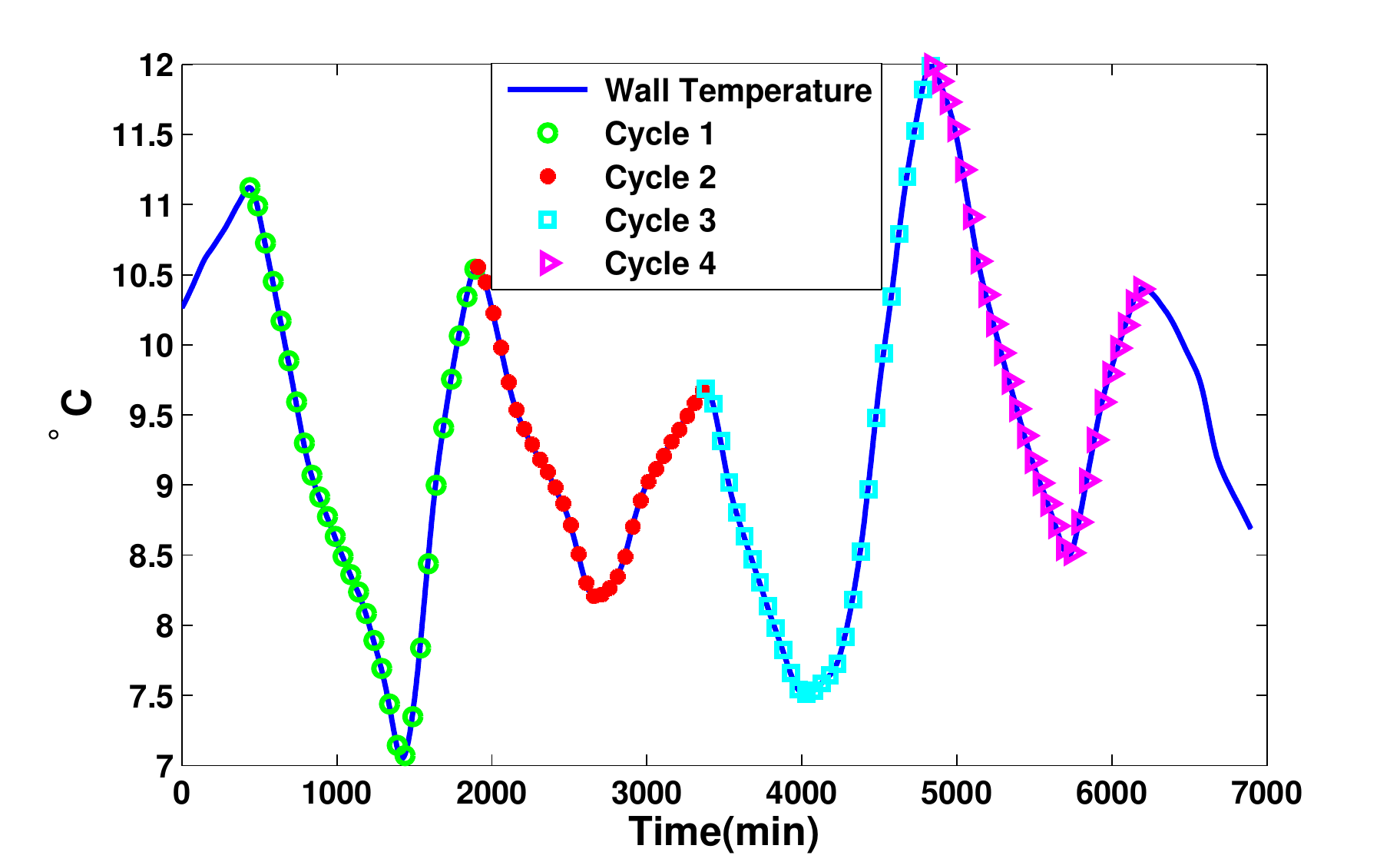}
\captionof{figure}{Different temperature cycles.}
\label{inf_gain_cycles}
\end{minipage}
\hfill
\begin{minipage}[b]{0.49\textwidth}
\centering
\begin{tabular}{|c|c|c|}
\hline
Cycle & Time (min)  &  $D_{KL}$ \\
\hline
1 & 1470 & 14.46 \\
\hline
2 & 1470 &  13.68 \\
\hline
3 & 1460  &  14.47 \\
\hline
4 & 1370  &  13.98 \\
\hline
\end{tabular} 
\captionof{table}{The estimated information gain from the detected external temperature oscillation cycles.}
\label{infgain_cycles}
\end{minipage}
\end{minipage}

\section{Summary and conclusions}

Our goal is to advance the mathematical modeling of the thermal properties of walls through statistical inference when in-situ measurements of surface temperatures and heat flux are available. Existing approaches are based on simplified models \cite{bidd}. Our approach uses the heat equation as a forward model to describe the heat dynamics of the wall. Moreover, our statistical methodology uses a marginalization technique that includes uncertainty in the temperature measurements, which are often incorporated as exact readings. We are thus able to achieve high error reduction and remove the bias due to the measurement error in the boundary temperature readings. Experimental data collected under controlled conditions demonstrate the utility of our method. To apply our methodology, we consider preprocessing techniques that are carefully chosen to fulfill convenient model assumptions. Consequently, we reduce the noise inherent in the experimental data and simplify the computations.

We considered two main numerical examples. In the first example, we explored a deterministic approach in which smoothed temperature measurements are used as exact boundary conditions. In the second example, we removed this deterministic constraint by modeling the nuisance boundary conditions as Gaussian random functions that are then marginalized analytically. In both examples, we derived the ML estimates of the thermal properties of the wall and the approximated posterior densities of such parameters by the Laplace method and the random-walk Metropolis-Hastings algorithm. We remark that, in both examples, the two techniques for the approximated Bayesian inference produce similar marginal posterior densities for the physical parameters of the wall. Based on that, one can use the Laplace method to obtain fast and efficient results and avoid costly MCMC simulations. We emphasize that the numerical results show that the utilization of our marginalization technique considerably reduces the bias error of the estimates of the model parameters, in contrast to when boundary conditions are assumed to be deterministic. Moreover, our estimates of the thermal resistance and the heat capacitance of the wall are consistent with the corresponding tabulated values for the wall under examination \cite{cibse2015environmental}. We carried out computationally intensive experiments to analyze the convergence of the MAP estimates, which are computed with the numerical solver when the space step, $\Delta x$, and the time step, $\Delta t$, are small. Besides, we have checked the robustness of our Bayesian inference in estimating parameters of the heat equation model, using the design and the application of subsampling algorithms with different subinterval lengths and subsample sizes.

Finally, in the Bayesian framework, we use the Laplace method again to estimate the information gain when the experimental setup consists either of the duration of the measurement campaign or the amplitude of the external temperature oscillation cycle. In this way, we can make recommendations on how to plan an efficient experimental design. Our numerical results indicate that a period of about 3.5 days is sufficient to gather data that allow the physical parameters of interest to be inferred with a high degree of accuracy. Moreover, our analysis shows that data corresponding to the oscillation cycles, which are characterized by a considerable variation in the temperature range, are informative. Our approach allows us to determine the optimal duration of the measurement campaign and the temperature oscillation cycle, both of which are valuable to practitioners. On the contrary, standard methods have required that data should be collected for two weeks, during the winter, such that large oscillations in temperature be avoided \cite{ISO869:2014}. The numerical examples reported in this work, applied to experimental data, indicate that our approach provides an accurate and robust methodology for inferring the thermal properties of solid brick walls, as well as for determining optimal experimental conditions for cost-effective measurement campaigns.

\section*{Acknowledgements}
Part of this work was carried out while M. Iglesias and M. Scavino were Visiting Professors at KAUST. \\ Z. Sawlan, M. Scavino and R. Tempone are members of the KAUST SRI Center for Uncertainty Quantification in Computational Science and Engineering. R. Tempone received support from the KAUST CRG3 Award Ref: 2281.

\section*{References}

\bibliography{mybib_v3}
\section{Appendix}

\subsection{Numerical approximation of the heat flux} \label{appA}

Let us introduce the following notation: $T_{m,n} = T(m \Delta x, n \Delta t), T_{int} (n\Delta t) = T_{int, n}$ and $T_{ext} (n\Delta t) = T_{ext, n}$. The backward Euler discretization of the heat equation in the interval, $(n \Delta t, (n+1) \Delta t)$, is given by

\begin{equation}
 \left\{
\begin{array}{rl}
\frac{1}{\Delta t} (T_{m, n+1} - T_{m , n}) & - \frac{\eta}{\Delta x ^2} (T_{m+1, n+1} -2 T_{m, n+1} + T_{m-1, n+1}) = 0,  \:\:m = 1,\ldots,M-1 \\
T_{0, n+1} =  & T_{int, n+1} \, , \\
T_{M, n+1} =  & T_{ext, n+1} \, ,
\end{array} \right.
\label{backeuler}
\end{equation}
where $\eta = \frac{k}{\rho C}$. \\

Let us consider the vectors $\mathop{\mathbf{T}_{n}}\limits_{(M-1) \times 1} = ( T_{1,n}, \ldots , T_{M-1,n})'$, $n = 0,\ldots,N$, and the matrix 
\begin{equation*}
\mathop{\mathbf{A}}\limits_{(M-1) \times (M-1)} = \left(
\begin{array}{cccccc}
-2 & 1 & 0 & 0 &  \ldots & 0 \\
1 & -2 & 1 & 0 & \ldots & 0 \\
0 & 1 & -2 & 1 & \ldots & 0 \\
\vdots & \ddots & \ddots & \ddots & \ddots & \vdots \\
0 & \ldots & 0 & 1 & -2 & 1 \\
0 & 0 & 0 & \ldots & 1 & -2
\end{array} \right) \,.
\end{equation*} 

The discretized system \eqref{backeuler} can be written in a matrix form as

\begin{equation} 
\frac{1}{\Delta t} (\mathbf{T}_{n+1} - \mathbf{T}_{n}) - \frac{\eta}{\Delta x^2} \mathbf{A} \mathbf{T}_{n+1}
= \frac{\eta}{\Delta x^2} (T_{int,n+1} \mathbf{a} + T_{ext,n+1} \mathbf{b}) \,, \label{eulermat} 
\end{equation}

where $\mathop{\mathbf{a}}\limits_{(M-1) \times 1} = (1, 0, \ldots, 0)'$, $\mathop{\mathbf{b}}\limits_{(M-1) \times 1} = (0, \ldots, 0, 1)'$.

The expression \eqref{eulermat} is equal to
\[ (I_{M-1} - \eta \frac{\Delta t}{\Delta x^2} \mathbf{A} ) \mathbf{T}_{n+1} = \mathbf{T}_{n} + \eta \frac{\Delta t}{\Delta x^2}
(T_{int,n+1} \mathbf{a} + T_{ext,n+1} \mathbf{b}) \, . \]

By letting $\frac{\Delta t}{\Delta x^2} = \lambda$ and $\mathbf{B} = (I_{M-1} - \eta \lambda \mathbf{A})^{-1}$, we obtain
\[ \mathbf{T}_{n+1} = \mathbf{B} \, \mathbf{T}_{n} + \eta \lambda ( T_{int,n+1} \mathbf{B} \, \mathbf{a} + T_{ext,n+1} \mathbf{B} \, \mathbf{b} ) \,. \]

Applying recursively the previous relation, we derive
\[ \mathbf{T}_{n} = \mathbf{B}^n \mathbf{T}_0 + \eta \lambda \sum_{k=1}^{n} T_{int,k} \mathbf{B}^{n-k+1} \mathbf{a} + \eta \lambda \sum_{k=1}^{n} T_{ext,k} \mathbf{B}^{n-k+1} \mathbf{b} \,. \]

Now, we approximate $F_{int}$ and $F_{ext}$ using forward and backward differences with second-order error:
\begin{eqnarray}
F_{int} (t_n) &\approx& \frac{k}{2\Delta x} \left( 3 T_{int,n} - 4 T_{1,n} + T_{2,n} \right), \label{fint} \\
F_{ext} (t_n) &\approx& \frac{k}{2\Delta x} \left( 3 T_{ext,n} -  4 T_{M-1, n} + T_{M-2, n} \right). \label{fext}
\end{eqnarray}

By defining the vectors $\mathop{\mathbf{c}}\limits_{(M-1) \times 1} = (-4, 1, 0, \ldots, 0)'$ and $\mathop{\mathbf{d}}\limits_{(M-1) \times 1} = (0, \ldots, 0, 1, -4)'$, we obtain 
\begin{eqnarray*}
F_{int} (t_n) &\approx& \frac{k}{2\Delta x} \left[ \mathbf{c}' \mathbf{B}^n \mathbf{T}_0 + 3  T_{int,n} + \eta \lambda \sum_{k=1}^{n} T_{int,k} \mathbf{c}' \mathbf{B}^{n-k+1} \mathbf{a} + \eta \lambda \sum_{k=1}^{n} T_{ext,k} \mathbf{c}' \mathbf{B}^{n-k+1} \mathbf{b}\right], \\
F_{ext} (t_n) &\approx& \frac{k}{2\Delta x} \left[ \mathbf{d}' \mathbf{B}^n \mathbf{T}_0 + \eta \lambda \sum_{k=1}^{n} T_{int,k} \mathbf{d}' \mathbf{B}^{n-k+1} \mathbf{a} + 3  T_{ext,n} + \eta \lambda \sum_{k=1}^{n} T_{ext,k} \mathbf{d}' \mathbf{B}^{n-k+1} \mathbf{b}\right] .
\end{eqnarray*}

Finally, we construct the matrices $\mathop{H}\limits_{(N+1) \times (M-1)}$, $\mathop{H_{int}}\limits_{(N+1) \times (N+1)}$ and $\mathop{H_{ext}}\limits_{(N+1) \times (N+1)}$ as follows:

\begin{itemize}
\item the matrix $H$ has the row vectors $ H^{i} = \frac{k}{2\Delta x} \mathbf{c}' \mathbf{B}^{i-1}, \, i = 1, \ldots, N+1$;

\item the matrix $H_{int}$ is lower triangular and is given by
\[  \frac{k \eta \lambda}{2\Delta x} \begin{bmatrix}
    \frac{3}{\eta \lambda} & 0 & 0 & \dots  & 0 \\
    0 & \frac{3}{\eta \lambda} + \mathbf{c}' \mathbf{B} \mathbf{a} & 0 & \dots  & 0 \\
   0 & \mathbf{c}' \mathbf{B}^2 \mathbf{a} &  \frac{3}{\eta \lambda} + \mathbf{c}' \mathbf{B} \mathbf{a} & \dots  & 0 \\
    \vdots & \vdots & \vdots & \ddots & \vdots \\
    0 &  \mathbf{c}' \mathbf{B}^{N} \mathbf{a} & \mathbf{c}' \mathbf{B}^{N-1} \mathbf{a} & \dots  & \frac{3}{\eta \lambda} + \mathbf{c}' \mathbf{B} \mathbf{a}
\end{bmatrix} \,;\]

\item the matrix $H_{ext}$  is lower triangular and is given by
\[  \frac{k \eta \lambda}{2\Delta x} \begin{bmatrix}
    0 & 0 & 0 & \dots  & 0 \\
    0 & \mathbf{c}' \mathbf{B} \mathbf{b} & 0 & \dots  & 0 \\
   0 & \mathbf{c}' \mathbf{B}^2 \mathbf{b} &  \frac{3}{\eta \lambda} + \mathbf{c}' \mathbf{B} \mathbf{b} & \dots  & 0 \\
    \vdots & \vdots & \vdots & \ddots & \vdots \\
    0 &  \mathbf{c}' \mathbf{B}^{N} \mathbf{b} & \mathbf{c}' \mathbf{B}^{N-1} \mathbf{b} & \dots  & \mathbf{c}' \mathbf{B} \mathbf{b} \end{bmatrix} \,. \]
\end{itemize}

Similarly, we can construct the matrices $\mathop{G}\limits_{(N+1) \times (M-1)}$, $\mathop{G_{int}}\limits_{(N+1) \times (N+1)}$ and $\mathop{G_{ext}}\limits_{(N+1) \times (N+1)}$.


\subsection{Marginal Likelihood}
\label{appB}

From equations (6),(7) and (8), the joint likelihood kernel of $\theta$ is given by

\small
\begin{equation*}
\begin{aligned}
\exp \big\{ & - \frac{1}{2} \left( \mathbf{Q}'_{int} \Sigma^{-1}_{int} \mathbf{Q}_{int} + (H\mathbf{T}_{0})' \Sigma^{-1}_{int} (H\mathbf{T}_{0})
- 2\mathbf{Q}'_{int}  \Sigma^{-1}_{int} (H\mathbf{T}_{0}) + \mathbf{Q}'_{ext} \Sigma^{-1}_{ext} \mathbf{Q}_{ext} + (G\mathbf{T}_{0})' \Sigma^{-1}_{ext} (G\mathbf{T}_{0}) - 2\mathbf{Q}'_{ext} \Sigma^{-1}_{ext} (G\mathbf{T}_{0}) \right)  \\ 
& - \frac{1}{2} \left( (H_{int}\mathbf{T}_{int})' \Sigma^{-1}_{int} (H_{int}\mathbf{T}_{int}) + 2(H_{int}\mathbf{T}_{int})' \Sigma^{-1}_{int} (H\mathbf{T}_{0}) + 2 (H_{int}\mathbf{T}_{int})' \Sigma^{-1}_{int} (H_{ext}\mathbf{T}_{ext}) 
-2 \mathbf{Q}'_{int} \Sigma^{-1}_{int} (H_{int}\mathbf{T}_{int}) \right) \\
& - \frac{1}{2} \left( (G_{int}\mathbf{T}_{int})' \Sigma^{-1}_{ext} (G_{int}\mathbf{T}_{int}) + 2(G_{int}\mathbf{T}_{int})' \Sigma^{-1}_{ext} (G\mathbf{T}_{0}) + 2 (G_{int}\mathbf{T}_{int})' \Sigma^{-1}_{ext} (G_{ext}\mathbf{T}_{ext}) 
-2 \mathbf{Q}'_{ext} \Sigma^{-1}_{ext} (G_{int}\mathbf{T}_{int}) \right) \\
& - \frac{1}{2} \left( (H_{ext}\mathbf{T}_{ext})' \Sigma^{-1}_{int} (H_{ext}\mathbf{T}_{ext}) + 2(H_{ext}\mathbf{T}_{ext})' \Sigma^{-1}_{int} (H\mathbf{T}_{0}) - 2 \mathbf{Q}'_{int} \Sigma^{-1}_{int} (H_{ext}\mathbf{T}_{ext}) \right) \\
& - \frac{1}{2} \left( (G_{ext}\mathbf{T}_{ext})' \Sigma^{-1}_{ext} (G_{ext}\mathbf{T}_{ext}) + 2(G_{ext}\mathbf{T}_{ext})' \Sigma^{-1}_{ext} (G\mathbf{T}_{0}) - 2 \mathbf{Q}'_{ext} \Sigma^{-1}_{ext} (G_{ext}\mathbf{T}_{ext}) \right) \\
& - \frac{1}{2} \left( \boldsymbol{\mu}'_{int} C^{-1}_{int,p} \boldsymbol{\mu}_{int} -2 \boldsymbol{\mu}'_{int} C^{-1}_{int,p} \mathbf{T}_{int} + \mathbf{T}'_{int} C^{-1}_{int,p} \mathbf{T}_{int} + \boldsymbol{\mu}'_{ext} C^{-1}_{ext,p} \boldsymbol{\mu}_{ext} -2 \boldsymbol{\mu}'_{ext} C^{-1}_{ext,p} \mathbf{T}_{ext} + \mathbf{T}'_{ext} C^{-1}_{ext,p} \mathbf{T}_{ext} \right) \big\} \, .
\end{aligned}
\end{equation*}

\normalsize

Let $U$ include any term that is independent from $\mathbf{T}_{int}$ and $\mathbf{T}_{ext}$; that is:
\begin{equation*}
\begin{aligned}
U = & \mathbf{Q}'_{int} \Sigma^{-1}_{int} \mathbf{Q}_{int} + \mathbf{Q}'_{ext} \Sigma^{-1}_{ext} \mathbf{Q}_{ext} + (H\mathbf{T}_{0})' \Sigma^{-1}_{int} (H\mathbf{T}_{0}) + (G\mathbf{T}_{0})' \Sigma^{-1}_{ext} (G\mathbf{T}_{0}) \\
& - 2 \mathbf{Q}'_{int} \Sigma^{-1}_{int} (H\mathbf{T}_{0}) - 2 \mathbf{Q}'_{ext} \Sigma^{-1}_{ext} (G\mathbf{T}_{0}) + \boldsymbol{\mu}'_{int} C^{-1}_{int,p} \boldsymbol{\mu}_{int} + \boldsymbol{\mu}'_{ext}C^{-1}_{ext,p} \boldsymbol{\mu}_{ext} \,.
\end{aligned}
\end{equation*}
Define
\begin{equation*}
\begin{aligned}
t'_{int,1} &= \left[ \mathbf{Q}'_{int} - (H\mathbf{T}_{0})' - (H_{ext}\mathbf{T}_{ext})' \right] \Sigma^{-1}_{int} H_{int} + \left[ \mathbf{Q}'_{ext} - (G\mathbf{T}_{0})' - (G_{ext}\mathbf{T}_{ext})' \right] \Sigma^{-1}_{ext} G_{int} + \boldsymbol{\mu}'_{int} C^{-1}_{int,p} \,, \\
\Lambda_0 &= \left( H'_{int} \Sigma^{-1}_{int} H_{int} + G'_{int} \Sigma^{-1}_{ext} G_{int} +  C^{-1}_{int,p} \right) ^{-1} \,.
\end{aligned}
\end{equation*}

By integrating first with respect to $\mathbf{T}_{int}$, the marginal likelihood of $\theta$ and $\mathbf{T}_{ext}$ is proportional to the product of a factor that is independent of $\mathbf{T}_{int}$ and the term
$(2 \pi)^{N/2} |\Lambda_0|^{1/2} \exp \left\{ \frac{1}{2} t'_{int,1} \Lambda_0 t_{int,1} \right\}\,.$

Now, let 
\begin{equation*}
\begin{aligned}
t'_{int,2} = & \left( \mathbf{Q}'_{int} - (H\mathbf{T}_{0})' \right) \Sigma^{-1}_{int} H_{int} + \left( \mathbf{Q}'_{ext} - (G\mathbf{T}_{0})' \right) \Sigma^{-1}_{ext} G_{int} + \boldsymbol{\mu}'_{int} C^{-1}_{int,p}, \\
\Lambda_1^{-1} = & H'_{ext} \Sigma^{-1}_{int} H_{ext} + G'_{ext} \Sigma^{-1}_{ext} G_{ext} +  C^{-1}_{ext,p} - 
(H'_{ext} \Sigma^{-1}_{int} H_{int} + G'_{ext} \Sigma^{-1}_{ext} G_{int}) \Lambda_0 ( H'_{int} \Sigma^{-1}_{int} H_{ext} + G'_{int} \Sigma^{-1}_{ext} G_{ext} )\, \\
t'_{ext,1} =& \left( \mathbf{Q}'_{int} - (H\mathbf{T}_{0})' \right) \Sigma^{-1}_{int} H_{ext} + \left( \mathbf{Q}'_{ext} - (G\mathbf{T}_0)' \right) \Sigma^{-1}_{ext} G_{ext} + \boldsymbol{\mu}'_{ext} C^{-1}_{ext,p} - t'_{int,2} \Lambda_0 ( H'_{int} \Sigma^{-1}_{int} H_{ext} + G'_{int} \Sigma^{-1}_{ext} G_{ext} ).
\end{aligned}
\end{equation*}

By integrating with respect to $\mathbf{T}_{ext}$, the marginal likelihood of $\theta$ is proportional to the product of a factor that is independent of $\mathbf{T}_{ext}$ and the term
$(2 \pi)^{N/2} |\Lambda_1|^{1/2} \exp \left\{ \frac{1}{2} t'_{ext,1} \Lambda_1 t_{ext,1} \right\}\,.$

\end{document}